\documentclass[showpacs,preprint,preprintnumbers,showpacs,showkeys,superscriptaddress,amsmath,amssymb,floatfix,superscriptaddress,nofootinbib]{revtex4}
\usepackage{bbold}
\usepackage{color}
\usepackage{latexsym}
\usepackage{amsmath}
\usepackage{amssymb}
\usepackage[utf8]{inputenc}
\usepackage{amsmath}
\usepackage{dsfont}
\usepackage{amsfonts}
\usepackage{bm}
\usepackage{bbold}
\usepackage{amssymb}
\usepackage{epstopdf}
\usepackage{epsfig}
\usepackage{eufrak}
\usepackage{euscript}
\usepackage{pstricks}
\usepackage{graphics}
\usepackage{graphicx}
\usepackage{picture}
\usepackage{appendix}
\usepackage{enumitem}
\usepackage{hyperref}

\newcommand{\be}{\begin{equation}}
\newcommand{\ee}{\end{equation}}
\newcommand{\ba}{\begin{eqnarray}}
\newcommand{\ea}{\end{eqnarray}}

\begin{document}

\title{\Large{Aspects of the quantization of non-linear electrodynamics in an uniform magnetic background}}
%

%\author{ J. M. A. Paix\~ao} \email{jeff@cbpf.br}
%\affiliation{Centro Brasileiro de Pesquisas F\'isicas, Rua Dr. Xavier Sigaud
%150, Urca, Rio de Janeiro, Brazil, CEP 22290-180}

%\author{ L. P. R. Ospedal } \email{leoopr@cbpf.br}
%\affiliation{Centro Brasileiro de Pesquisas F\'isicas, Rua Dr. Xavier Sigaud
%150, Urca, Rio de Janeiro, Brazil, CEP 22290-180}

\author{M. J. Neves} \email{mariojr@ufrrj.br}
%\affiliation{Department of Physics and Astronomy, University of Alabama, Tuscaloosa, Alabama 35487, USA}
\affiliation{Departamento de F\'isica, Universidade Federal Rural do Rio de Janeiro, BR 465-07, 23890-971, Serop\'edica, RJ, Brazil.}

%\author{J. A. Hela\"yel-Neto}\email{helayel@cbpf.br}
%\affiliation{Centro Brasileiro de Pesquisas F\'isicas, Rua Dr. Xavier Sigaud
%150, Urca, Rio de Janeiro, Brazil, CEP 22290-180}
%

%\date{\today}

%\pacs{04.50.-h, 05.20.-y, 05.90.+m}

%\arxivnumber{....arxiv:}

\begin{abstract}
\noindent

The canonical and path integral quantization for non-linear electrodynamics in the presence of an external electromagnetic (EM) field 
are proposed in this work. The EM background field is introduced expanding a general lagrangian of a generic non-linear 
electrodynamics around the background for small fluctuations of the propagating fields, where we consider up to the quadratic terms 
in the propagating fields. As consequence, we obtain an electrodynamics linearized by the presence of the external EM field. Thereby, we study the 
canonical quantization calculating the energy for the ground state of the linearized EM field in terms of an external magnetic field. 
The microcausality of the model also is discussed through the Pauli-Jordan function. We also define a generating functional for the 
linearized EM field, in which, in the Coulomb gauge, the Green function of the model is obtained, and we can construct the perturbative formalism 
like in standard quantum field theory (QFT). As application of the perturbation theory, the effective potential at one loop is calculated for the 
linearized EM field coupled to a complex scalar field. We apply the results in the case of the Modified Maxwell ED.

%\clearpage
%
\end{abstract}

%\pacs{11.15.-q,12.60.-i,11.10.Lm}
%
%\keywords{Non-linear quantum electrodynamics, electron's anomalous momentum.
%}

\maketitle

\newpage

\pagestyle{myheadings}
\markright{Aspects of the quantization of NLEDs in an uniform magnetic background}
%
%%%%%%%%%%%%%%%%%%%%%%%%%%%%%%%
%
%%%%%%%%%%%%%%%%%%%%%%%%%%%%%%%
%
%\newcommand{\be}{\begin{equation}}
%\newcommand{\ee}{\end{equation}}
%\newcommand{\ba}{\begin{eqnarray}}
%\newcommand{\ea}{\end{eqnarray}}
%\newcommand{\p}{\partial}
%\def\ni{\noindent}
%\def\Dc{{\cal D}}
%\def\no{\nonumber}
%
%%%%%%%%%%%%%%%%%%%%%%%%%%%%%%%%%%%%%%%%%%%%%%%%%%%%%%%%%%%%%%%%%%%%%%%%%%%%%%%%%%%%%%%%%%%%%%%%%%%%%%%%%%%%%%%%%%%%%%%%%%%
%
%\renewcommand{\theequation}{1.\arabic{equation}}
%\setcounter{equation}{0}
%
%\newpage
%
%\section{Introduction}
%\renewcommand{\theequation}{1.\arabic{equation}}
%\setcounter{equation}{0}

%%%%%%%%%%%%%%%%
\section{Introduction}
%%%%%%%%%%%%%%%%
\label{sec:1}
Non-linear electrodynamics (NLEDs) are extensions of the Maxwell electrodynamics (MED) in which electric and magnetic fields 
emerges through the higher order terms in the correspondent lagrangian. Since it is well known in the literature, the first NLED was 
proposed by M. Born and L. Infeld to explain the electrostatic field of a like-point charge at origin \cite{BornI}. In parallel, the 
quantum electrodynamics (QED) was formulated to understand the interaction of photons with electrons and positrons \cite{Fermi1,Fock,Fermi2,Dirac}.
In 1936, H. Euler and W. Heisenberg calculated the effective action for fermions coupled to EM field in the QED, obtaining the NL 
Euler-Heisenberg ED that described with excellent prediction the light-by-light scattering \cite{HEuler}. One of the optical effects 
that motivate the study of NLEDs is the vacuum birefringence, when the quantum vacuum is stressed by external electromagnetic fields 
in which it behaves like a material medium. Explicitly, the birefringence correspond to variation of the refractive index associated with a 
material medium, when the wave polarization that crosses the medium changes with the external magnetic field. This effect was studied in different contexts of NLEDs, 
see the refs. \cite{Adler,Pistoni,Ruffini,Dunne,Rizzo,Sarazin,Russo2022,Bamber,Burke,Tommasini}. However, it has not been confirmed yet \cite{Ejlli}. 
Nowadays, NLEDs also arise in several research areas. The Born-Infeld effective actions emerge from superstrings theories, alternative gravitation models, 
black holes and theories with magnetic monopoles \cite{Fradkin,Pope,Banerjee,Mann,Barrientos22,Ali22,Garcia}. The scenario of a NLED in the presence of external EM field opens the possibility of investigations in the properties of wave propagation, as dispersions relations, group velocities \cite{MJNevesPRD2021}. The coupling of axions with a general NLED has been proposed as a way to obtain the birefringence when the model is submitted to external electromagnetic fields \cite{nossoJHEP1,nossoJHEP2}.  In material physics, loop calculations from modified QED in Dirac materials imply into a type-Euler-Heisenberg electrodynamics for weak electric and magnetic fields \cite{Keser}. For strong EM fields, a new NLED candidate to description of Dirac materials is proposed in the ref. \cite{NevesJPA}. In connection with the particle physics, the ATLAS collaboration
imposed the limit of $90$ GeV for the Born-Infeld scale via light-by-light scattering in LHC Pb-Pb collisions \cite{Ellis}. Extensions of the Standard Model (SM) with the $U(1)_{Y}$ hypercharge group associated with Born-Infeld was used to generate anomalous self-couplings between photon and $Z$-boson constraining decay processes beyond the SM \cite{Gaete_AHEP_2021,Gaete_EPJC_2022}. Other types of NLEDs with a logarithm function also were proposed \cite{GENBI,LOG}. For review and notes on NLEDs, see the refs. \cite{Plebanski68,Birula70,Boillat70,Boillat72,Bialynicki83,Paixao}.   
A new modification of the MED was proposed few years ago that preserves all the symmetries, including duality and the conformal symmetry, and also preserves important proprieties of a QFT, as causality and unitarity. This theory is known as modified Maxwell electrodynamics (ModMax ED) \cite{Sorokin1,Sorokin2,Sorokin3}. 
This NLED has attracted the attention due to contributions in light-by-light scattering, novel black holes solutions, $T\,\bar{T}$-like deformations of the Maxwell ED, in context of string theories, and supersymmetry \cite{Maceda21,Bordo21,Amirabi21,Kruglov2022,Velni22,Yekta,Conti22,Ferko22,Nastase2022,Denisov17,SusyModmax}. 
Under both electric and magnetic external fields, the ModMax ED shows that the vacuum has birefringence \cite{NevesPRD2023}, and others optical phenomena as Goos-H\"anchen effect and complex Kerr rotation 
\cite{NevesEPJPlus}.

All the motivations lead us to investigations of the quantization of NLEDs that are linearized by the presence of an external EM field, that we consider uniform and constant in this work. We start with the canonical quantization of an any linearized model in which the quantized energy is affected by the external magnetic field. The Pauli-Jordan function is obtained in terms of the dispersion relations of the NLEDs. We introduce the path integral quantization defining the generating functional for the linearized ED fixing a Coulomb gauge. In this gauge, we obtain the Green function in the presence of the external magnetic field. The functional approach is constructed to calculate the effective potential of a complex scalar field coupled to the linearized gauge field. Since the results are valid in any NLED, we use the ModMax ED as example of application, exploring the vacuum energy, microcausality and the effects of the ModMax parameter in the effective potential. 
The paper is organized as follows : In the section (\ref{sec2}), we show a review on the NLEDs in external EM fields. The section (\ref{sec3}) is presented the canonical quantization. In the section (\ref{sec4}), the path integral quantization is discussed. The section (\ref{sec5}) is due to formulation of the effective potential in a linearized scalar ED. The effective potential is renormalized at one loop in the section (\ref{sec6}). In the section (\ref{sec7}), we apply the results of the previous sections for the ModMax ED. For end, the conclusions are presented in the section (\ref{sec8}).   
In this paper, we use the natural units of $\hbar=c=1$ in which the Minkowski metric has the signature $\eta^{\mu\nu}=\mbox{diag}(+1,-1,-1,-1)$.
%

%
%%%%%%%%%%%%%%%%%%%%%%%%%%%
\section{Review on non-linear electrodynamics in a electromagnetic background field}
%%%%%%%%%%%%%%%%%%%%%%%%%%%
\label{sec2}
In this section, we show a brief review of an NLED in a EM background field.
We start up with a  
%Lagrangian of the model
%%
%\begin{eqnarray}\label{Lmodel}
%{\cal L}={\cal L}_{nl}({\cal F},{\cal G})
%\,+\,\overline{\psi} \, \left( \, i \, \gamma^{\mu}D_{\mu} - m \, \mathds{1} \, \right) \psi
%+\frac{1}{2} \, \left(\partial_{\mu}\phi \right)^{2}
%-\frac{1}{2} \, m^2 \, \phi^2
%+ g \, \phi \, {\cal G}-J_{\mu}\,A^{\mu}
%\, \; ,
%\end{eqnarray}
%
general Lagrangian density 
\begin{eqnarray}\label{Linitial}
\mathcal{L}=\mathcal{L}_{nl}({\cal F}_{0},{\cal G}_{0})-J_{\mu}\,A_{0}^{\;\,\mu} \; ,
\end{eqnarray}
that is function of the Lorentz and gauge invariants
$
{\cal F}_{0}=-\frac{1}{4} \, F_{0\mu\nu}^{2}=\frac{1}{2} \, \left( {\bf E}_{0}^2-{\bf B}_{0}^2\right),
$
and
$
{\cal G}_{0}=-\frac{1}{4} \, F_{0\mu\nu}\widetilde{F}_{0}^{\;\mu\nu}={\bf E}_{0}\cdot{\bf B}_{0}
$, 
%$\psi$ sets a fermion field, $D_{\mu}=\partial_{\mu}+i \, e \, A_{\mu}$ is the
%abelian covariant derivative operator, $m$ is the fermion mass and $\gamma^{\mu}$ are the Dirac matrices,
%that satisfy the relation $\gamma^{\mu} \, \gamma^{\nu}=\eta^{\mu\nu} \, \mathds{1}-i \, \Sigma^{\mu\nu}$.
%$\left\{ \, \gamma^{\mu} \, , \, \gamma^{\nu} \, \right\}=2\,\eta^{\mu\nu} \, \mathbb{1}$ and
%$\sigma^{\mu\nu}=i\left[\,\gamma^{\mu}\,,\,\gamma^{\nu}\,\right]/2$.
%The Lagrangian (\ref{Lmodel}) is clearly $U(1)$ gauge invariant.
in which the EM strength field tensor has the components $F_{0}^{\mu\nu}=\partial^{\mu}A_{0}^{\nu}-\partial^{\nu}A_{0}^{\mu}=\left( \, -E_{0}^{i} \, , \, -\epsilon^{ijk}B_{0}^{k} \, \right)$, and the correspondent dual tensor is $\widetilde{F}_{0}^{\mu\nu}=\epsilon^{\mu\nu\alpha\beta}\,F_{0\alpha\beta}/2=\left( \, -B_{0}^{i} \, , \, \epsilon^{ijk}E_{0}^{k} \, \right)$. The dual tensor satisfies the Bianchi identity $\partial_{\mu}\widetilde{F}_{0}^{\;\mu\nu}=0$. In (\ref{Linitial}), 
$J^{\mu}$ is a conserved $4$-current that satisfies the continuity equation $\partial_{\mu}J^{\mu}=0$.

The $A_{0}^{\mu}$-potential is written as $A_{0}^{\mu}(x)=a^{\mu}(x)+A_{B}^{\;\;\,\,\mu}(x)$, in which we define $a^{\mu}$ as the propagating gauge field,
and $A_{B}^{\mu}(x)$ is the potential associated with the EM background field. In this conjecture, the expansion implies into writing the EM strength tensor 
as the combination $F_{0}^{\;\mu\nu}(x)=f^{\mu\nu}(x) \, + \, F_{B}^{\;\;\,\mu\nu}(x)$, where $f^{\mu\nu}=\partial^{\mu}a^{\nu}-\partial^{\nu}a^{\mu}=\left( \, -e^{i} \, , \, -\epsilon^{ijk}b^{k} \, \right)$ 
is the field strength tensor related with the propagating gauge field, and $F_{B}^{\;\;\,\mu\nu}=\left( \, -E^{i} \, , \, -\epsilon^{ijk} \, B^{k} \, \right)$ corresponds to the EM background field. 
In general, the electric ${\bf E}$ and magnetic ${\bf B}$ depends on the space-time coordinates.
%
%null the electric background, {\it i.e.},
%$E^{i}=0$, and the magnetic background is constant and uniform.
%
Under these general conditions, we expand the Lagrangian $\mathcal{L}_{nl}({\cal F}_{0},{\cal G}_{0})$ 
up to second order in the perturbation theory around the strength field tensor $f^{\mu\nu}$ (for small fluctuations of propagating gauge field) 
to obtain the expression \cite{MJNevesPRD2021}
\begin{equation}\label{L2}
{\cal L}_{nl}^{(2)}=-\frac{1}{4} \, c_{1} \, f_{\mu\nu}^{\, 2}
-\frac{1}{4} \, c_{2} \, f_{\mu\nu}\widetilde{f}^{\mu\nu}
-\frac{1}{2} \, f_{\mu\nu} G_{B}^{\;\;\,\mu\nu}+\frac{1}{8} \, Q_{B}^{\; \; \; \mu\nu\kappa\lambda}f_{\mu\nu} \, f_{\kappa\lambda}
%+\frac{1}{8} \, Q_{B}^{\; \; \; \mu\nu\kappa\lambda}f_{\mu\nu}\widetilde{f}_{\kappa\lambda}
%+\frac{1}{2} \, \left(\partial_{\mu}\phi \right)^{2}
%-\frac{1}{2} \, m^2 \, \phi^2
%\nonumber \\
%&&
%\hspace{-0.3cm}
%+\frac{1}{8} \, R_{B}^{ \; \; \; \mu\nu\kappa\lambda\rho\sigma}f_{\mu\nu}f_{\kappa\lambda}f_{\rho\sigma}
%+\frac{1}{16} \, S_{B}^{ \; \; \; \mu\nu\kappa\lambda\rho\sigma\omega\tau}f_{\mu\nu}f_{\kappa\lambda}f_{\rho\sigma}f_{\omega\tau}
%-\frac{1}{4} \, g \, \phi \, f_{\mu\nu}\widetilde{f}^{\mu\nu}
%- \frac{1}{2} \, g \, \phi \, f_{\mu\nu}\widetilde{F}_{B}^{\;\;\,\mu\nu}
%+ g \, \phi \, \matches{G}_{B}
%-J_{\mu}\,a^{\mu}-J_{\mu}A_{B}^{\mu}
%+\,\overline{\psi} \, \left( \, i \, \gamma^{\mu}\partial_{\mu}-e \, \gamma^{\mu}A_{B\mu} - m \, \mathds{1} \, \right) \psi
%-e \, \overline{\psi} \, \gamma^{\mu}a_{\mu} \, \psi
+{\cal L}_{nl}\left( {\cal F}_{B} , {\cal G}_{B} \right)-J_{\mu}(a^{\mu}+A_{B}^{\mu}) \; ,
\end{equation}
where the background tensors are defined by
\begin{subequations}
\begin{eqnarray}
G_{B}^{\;\;\;\,\mu\nu} &=& c_{1} \, F_{B}^{ \, \; \; \mu\nu}+c_{2} \, \widetilde{F}_{B}^{\, \; \; \mu\nu} \; ,
\label{GB}
\\
Q_{B}^{\; \; \; \mu\nu\kappa\lambda} &=& d_{1} \, F_{B}^{\; \; \, \mu\nu}F_{B}^{\; \; \, \kappa\lambda}
+d_{2} \, \widetilde{F}_{B}^{\, \; \; \mu\nu}\widetilde{F}_{B}^{\, \; \; \kappa\lambda}
+d_{3} \, F_{B}^{\, \; \; \mu\nu}\widetilde{F}_{B}^{\, \; \; \kappa\lambda}
+ d_{3} \, \widetilde{F}_{B}^{\, \; \; \mu\nu} F_{B}^{\, \; \; \kappa\lambda} \; ,
\label{QB}
\end{eqnarray}
\end{subequations}
and ${\cal L}_{nl}\left({\cal F}_{B},{\cal G}_{B}\right)$ is function only of the EM background field, with
${\cal F}_{B}=-\frac{1}{4} \, F_{B\mu\nu}^2={\bf E}^2-{\bf B}^2$ and ${\cal G}_{B}=-\frac{1}{4} \, F_{B\mu\nu}\widetilde{F}_{B}^{\;\;\,\mu\nu}={\bf E}\cdot{\bf B}$.
%, both in terms of the background electromagnetic field.
%The current $J^{\mu}$ couples to the external potential $A_{B}^{\mu}$ but this term is irrelevant for the field equations.
The background tensors, $G_{B}^{\;\;\;\,\mu\nu}=-G_{B}^{\;\;\;\,\nu\mu}$ and $Q_{B}^{\; \; \; \mu\nu\kappa\lambda}$, are symmetric under exchange 
$\mu\nu \leftrightarrow \kappa\lambda$, and antisymmetric under $\mu \leftrightarrow \nu$ and $\kappa \leftrightarrow \lambda$. The coefficients of the expansion 
$c_{1}$, $c_{2}$, $d_{1}$, $d_{2}$ and $d_{3}$ are evaluated at the EM background :
\begin{eqnarray}\label{coefficients}
c_{1}&=&\left.\frac{\partial{\cal L}_{nl}}{\partial{\cal F}}\right|_{B}
\; , \;
\left. c_{2}=\frac{\partial{\cal L}_{nl}}{\partial{\cal G}}\right|_{B}
\; , \;
\left. d_{1}=\frac{\partial^2{\cal L}_{nl}}{\partial{\cal F}^2}\right|_{B}
\; , \;
\left. d_{2}=\frac{\partial^2{\cal L}_{nl}}{\partial{\cal G}^2}\right|_{B}
\; , \;
\left. d_{3}=\frac{\partial^2{\cal L}_{nl}}{\partial{\cal F}\partial{\cal G}}\right|_{B} \; ,
\end{eqnarray}
that, in general, also are functions of the space-time coordinates.
If we consider the EM background field uniform and constant, $\partial_{\alpha}F_{B\mu\nu}=0$, and the $A_{B}^{\mu}$-potential is 
$A_{B}^{\mu}(x)=-F_{B}^{\,\, \mu\nu}\,x_{\nu}/2$, and the field strength tensor is reduced to $F_{0}^{\;\mu\nu}(x)=f^{\mu\nu}(x) \, + \, F_{B}^{\;\;\,\mu\nu}$. 
As consequence, the third term in (\ref{L2}) is a surface term inside the correspondent 
action and it can be discarded in the lagrangian. The last term ${\cal L}_{nl}\left( \, {\cal F}_{B} \, , \, {\cal G}_{B} \, \right)$ 
does not reproduce any dynamics for the system, and it can be absorbed in the final lagrangian. All the coefficients of (\ref{coefficients}) 
are constants and uniforms, it depends on the parameters that set the NLED in question. We also consider examples of NLEDs 
with CP-invariance in which, in the minimal, the Lagrangian density depends on the ${\cal G}^n$, for $n$ any even number. When we have only 
an uniform magnetic background field, or only an uniform electric background, the coefficients $c_2=d_3=0$ in all the examples of NLEDs known 
in the literature. Under all these conditions, substituting the tensor (\ref{QB}) in (\ref{L2}), the quadratic lagrangian is reduced to
\begin{eqnarray}\label{L2simpc1}
{\cal L}_{nl}^{(2)} =-\frac{1}{4} \, c_{1} \, f_{\mu\nu}^{\, 2}
%-\frac{1}{4} \, c_{2} \, f_{\mu\nu}\widetilde{f}^{\mu\nu}
%-\frac{1}{2} \, f_{\mu\nu} \, G_{B}^{\;\;\,\mu\nu}
+\frac{d_1}{8} \, (F_{B\mu\nu}f^{\mu\nu})^{2}
+\frac{d_2}{8} \, (\widetilde{F}_{B\mu\nu}f^{\mu\nu})^{2}-J_{\mu}\,a^{\mu} \; .
%+\frac{1}{8} \, Q_{B}^{\; \; \; \mu\nu\kappa\lambda}f_{\mu\nu}\widetilde{f}_{\kappa\lambda}
%+\frac{1}{2} \, \left(\partial_{\mu}\phi \right)^{2}
%-\frac{1}{2} \, m^2 \, \phi^2
%\nonumber \\
%&&
%\hspace{-0.3cm}
%+\frac{1}{8} \, R_{B}^{ \; \; \; \mu\nu\kappa\lambda\rho\sigma}f_{\mu\nu}f_{\kappa\lambda}f_{\rho\sigma}
%+\frac{1}{16} \, S_{B}^{ \; \; \; \mu\nu\kappa\lambda\rho\sigma\omega\tau}f_{\mu\nu}f_{\kappa\lambda}f_{\rho\sigma}f_{\omega\tau}
%-\frac{1}{4} \, g \, \phi \, f_{\mu\nu}\widetilde{f}^{\mu\nu}
%- \frac{1}{2} \, g \, \phi \, f_{\mu\nu}\widetilde{F}_{B}^{\;\;\,\mu\nu}
%+ g \, \phi \, \mathcal{G}_{B}
%-J_{\mu}\,a^{\mu}-J_{\mu}A_{B}^{\mu}
%+\,\overline{\psi} \, \left( \, i \, \gamma^{\mu}\partial_{\mu}-e \, \gamma^{\mu}A_{B\mu} - m \, \mathds{1} \, \right) \psi
%-e \, \overline{\psi} \, \gamma^{\mu}a_{\mu} \, \psi
%+{\cal L}_{nl}\left( \, {\cal F}_{B} \, , \, {\cal G}_{B} \, \right) \; ,
\end{eqnarray} 
The usual Maxwell ED is recovered in (\ref{L2simpc1}) when we fix $c_1=1$ and $d_{1}=d_{2}=0$. We can notice that the same limit 
is obtained when the background tensor is turned off. We can absorb the $c_1$-factor in the $4$-potential making $\sqrt{c_1}\,a^{\mu}\rightarrow a^{\mu}$, 
such that the lagrangian is rewritten as 
\begin{eqnarray}\label{L2simp}
{\cal L}_{nl}^{(2)} =-\frac{1}{4} \, f_{\mu\nu}^{\, 2}
+\frac{d_B}{8} \, (F_{B\mu\nu}f^{\mu\nu})^{2}
+\frac{d_E}{8} \, (\widetilde{F}_{B\mu\nu}f^{\mu\nu})^{2}-J_{\mu}\,a^{\mu} \; ,
\end{eqnarray} 
where $d_B=d_1/c_1$, $d_E=d_2/c_1$, and the source term is redefined as $\sqrt{c_1} \, J^{\mu} \rightarrow J^{\mu}$.
The action associated with the lagrangian (\ref{L2simp}) remains gauge invariant for the 
transformations $a^{\prime\mu}=a^{\mu}+\partial^{\mu}\Lambda$ for any scalar $\Lambda$-function, if the $4$-current is conserved, {\it i.e.}, $\partial_{\mu}J^{\mu}=0$.
For convenience, we add the gauge fixing term ${\cal L}_{gf}=-(\partial_{\mu}a^{\mu})^2/(2\xi)$ to the lagrangian (\ref{L2simp}) for a gauge fixing parameter $\xi$.
The action principle applied to the total lagrangian ${\cal L}^{(2)}+{\cal L}_{gf}$ yields the source equation
\begin{eqnarray}
\partial^{\mu}f_{\mu\nu}+\frac{1}{\xi}\,\partial_{\nu}\left(\partial_{\mu}a^{\mu}\right)-\frac{d_B}{2} \, F_{B\mu\nu}\,\partial^{\mu}(F_{B\kappa\lambda}f^{\kappa\lambda})
-\frac{d_E}{2} \, \widetilde{F}_{B\mu\nu}\,\partial^{\mu}(\widetilde{F}_{B\kappa\lambda}f^{\kappa\lambda}) = J_{\nu} \; ,
\end{eqnarray}
that in terms of $4$-potential $a^{\mu}$ is
\begin{eqnarray}\label{Eqanu}
\left[\,\eta_{\mu\nu}\Box+\left( \frac{1}{\xi}-1 \right)\partial_{\mu}\,\partial_{\nu} -\left(d_{B}\,F_{B\mu\alpha}F_{B\nu\beta}+d_{E}\,\widetilde{F}_{B\mu\alpha}\widetilde{F}_{B\nu\beta}\right)\partial^{\alpha}\partial^{\beta} \,  \right] a^{\mu}=J_{\nu} \; .
\end{eqnarray}
The solution of inhomogeneous equation (\ref{Eqanu}) is   
\begin{eqnarray}\label{solamu}
a_{\mu}(x)=a_{0\mu}(x)+\int d^{4}x^{\prime} \, \Delta_{\mu\nu}(x-x^{\prime})\,J^{\nu}(x^{\prime}) \; ,
\end{eqnarray}
where $a_{0\mu}(x)$ is solution of the homogeneous equation, and $\Delta_{\mu\nu}(x-x^{\prime})$ is the Green function of the operator 
in (\ref{Eqanu}) that satisfies the differential equation 
\begin{equation}\label{EqDelta}
\left[ \, \eta_{\mu\nu}\Box+\left( \frac{1}{\xi}-1 \right)\partial_{\mu} \, \partial_{\nu}
-\left(d_{B}\,F_{B\mu\alpha}F_{B\nu\beta}+d_{E}\,\widetilde{F}_{B\mu\alpha}\widetilde{F}_{B\nu\beta}\right)\partial^{\alpha}\partial^{\beta} \,\right]\Delta^{\nu}_{\;\;\rho}(x-x^{\prime})=\eta_{\mu\rho}\,\delta^{4}(x-x^{\prime}) \; .
\end{equation}
The Fourier transform for $\Delta_{\mu\nu}(x-x^{\prime})$ provides the inverse of the operator in (\ref{EqDelta}), in which we will discuss the conditions to obtain the inversion of (\ref{EqDelta}) in the section (\ref{sec4}).

\section{Canonical quantization of a linearized ED}
\label{sec3}
In this section, we show the canonical quantization in the Coulomb gauge of any NLED when it is submitted to an external magnetic background field. 
From free lagrangian (\ref{L2simp}), the correspondent canonical momentum is expressed by 
\begin{eqnarray}
\pi^{\nu}=\frac{\partial {\cal L}^{(2)}}{\partial \dot{a}_{\nu}}=-f^{0\nu}+\frac{d_B}{2} \left(F_{B\alpha\beta}f^{\alpha\beta}\right) F_{B}^{\;\;\,0\nu}
+\frac{d_E}{2} \left(\widetilde{F}_{B\alpha\beta}f^{\alpha\beta}\right) \widetilde{F}_{B}^{\;\;\,0\nu} \; ,
\end{eqnarray}
that for $F_{B}^{\;\;\,\mu\nu}=\left( \, 0 \, , \, -\epsilon^{ijk} \, B^{k} \, \right)$ and $\widetilde{F}_{B}^{\;\;\,\mu\nu}=\left( \, -B^{i} \, , \, 0 \, \right)$, the $\pi^{\nu}$-components are 
\begin{eqnarray}
\pi^{0}=0 
\quad  , \quad
\pi^{i}=e^{i}+d_{E} \left({\bf B} \cdot {\bf e}\right) B^{i} \; .
\end{eqnarray}
In Coulomb gauge, the potentials are constrained by the condition $a^{0}=\nabla\cdot{\bf a}=0$ in which the free equation (\ref{Eqanu}) is written as $\hat{{\cal O}}_{ij}\,a^{i}=0$, where the operator $\hat{{\cal O}}_{ij}$ is
\begin{eqnarray}\label{opOij}
\hat{{\cal O}}_{ij}=\eta_{ij}\,\Box-d_B\left[ \, \delta_{ij} ({\bf B}\times \nabla)^2+({\bf B} \cdot \nabla) B_{i}\,\partial_{j}-B_{i}\,B_{j}\,\nabla^2 \,\right]-d_E\, B_{i} \, B_{j} \, \partial_{t}^2 \; . 
\end{eqnarray}
 Using the plane wave solutions for $a^{i}$ with wave vector ${\bf k}$, frequency $\omega$ and $a_0^{\,i}$, the $a^{i}$-equation is reduced to  ${\cal O}_{ij}(\omega,{\bf k})\,a_{0}^{i}=0$, in which ${\cal O}_{ij}$ is the operator (\ref{opOij}) in the $({\bf k},\omega)$-space. The non-trivial solutions of this equation are for $\det ({\cal O}_{ij})=0$ that leads the dispersion relations \cite{MJNevesPRD2021} :
\begin{eqnarray}\label{omegask}
\omega_{1}({\bf k})=|{\bf k}| 
\; , \;
\omega_{2}({\bf k})=|{\bf k}| \, \sqrt{1-d_{B} \, ( \hat{{\bf k}} \times {\bf B} )^2 }
\; , \;
\omega_{3}({\bf k})=|{\bf k}| \, \sqrt{1-\frac{d_{E} \, ( \hat{{\bf k}} \times {\bf B} )^2}{1+d_{E}\,{\bf B}^2} } \; .
\end{eqnarray}
The first dispersion relation $\omega_{1}({\bf k})$ in (\ref{omegask}) is due to gauge invariance of the linearized ED. The other solutions $\omega_{2}({\bf k})$ and $\omega_{3}({\bf k})$ depend on the $d_{B}$- and $d_{E}$-coefficients. Substituting the plane waves solutions for the electric $({\bf e})$ and magnetic $({\bf b})$-fields in the free equations of linearized ED, the vectors ${\bf k}$ and ${\bf e}$ are perpendiculars if $d_{B} \neq 0$ and $d_{E}=0$, and consequently, the frequency
$\omega_{2}({\bf k})$ is a transversal mode. Considering a particular case in which $d_{B}=0$ and $d_{E} \neq 0$, the vectors ${\bf k}$ and ${\bf e}$ are not perpendiculars 
for a free plane wave, and as consequence, $\omega_{3}({\bf k})$ is not a transversal mode. For more details, see the ref. \cite{MJNevesPRD2021}.
The canonical quantization approach promotes $a^{i}$ and $\pi^{i}$ to the operators $\hat{a}^{i}$ and $\hat{\pi}^{i}$ 
that satisfy the commutation relations :
\begin{subequations}
\begin{eqnarray}
\left[ \,  \hat{a}^{i}({\bf x},t) \, , \,  \hat{a}^{j}({\bf x}^{\prime},t) \, \right] &=& \left[ \,  \hat{\pi}^{i}({\bf x},t) \, , \,  \hat{\pi}^{j}({\bf x}^{\prime},t) \, \right]  = 0 
\; , \; 
\\
\left[ \,  \hat{a}^{i}({\bf x},t) \, , \,  \hat{\pi}^{j}({\bf x}^{\prime},t) \, \right]  &=& i \, \left( \delta^{jk}+d_{E}\,B^{j}\,B^{k}  \right)  \hat{P}^{ik} \, \delta^{3}({\bf x} - {\bf x}^{\prime}) \; ,
\label{relapi}
\end{eqnarray}
\end{subequations}
where $\hat{P}^{ik}=\delta^{ik}-\partial^{i}\,\partial^{k}/\nabla^2$, and the last relation (\ref{relapi}) agrees with the Coulomb gauge condition. 
The general solution for the $\hat{a}^{i}$-operator is given by
\begin{eqnarray}\label{solaiop}
\hat{a}^{i}({\bf x},t)=\int \frac{d^3{\bf k}}{\sqrt{(2\pi)^{3} \, 2\,\omega({\bf k})}} \, \sum_{\lambda=1}^{2} \epsilon^{i}({\bf k},\lambda) \left[ \, \hat{a}^{(\lambda)}({\bf k}) \, e^{-ik \cdot x} + \hat{a}^{(\lambda)\dagger}({\bf k}) \, e^{ik \cdot x}  \, \right] \; ,
\end{eqnarray}
where $\omega({\bf k})$ is any frequency from (\ref{omegask}), $\epsilon^{i}({\bf k},\lambda)$ are the components of the polarization vector, the scalar product is defined by $k \cdot x=t\,\omega({\bf k})-{\bf k}\cdot{\bf x}$, and the annihilation $\hat{a}^{(\lambda)}({\bf k})$ and creation $\hat{a}^{(\lambda)\dagger}({\bf k})$ operators satisfy the commutation relations in the standard form
\begin{subequations}
\begin{eqnarray}
\left[ \,  \hat{a}^{(\lambda)}({\bf k}) \, , \,  \hat{a}^{(\lambda^{\prime})}({\bf k}^{\prime}) \, \right] &=& \left[ \, \hat{a}^{(\lambda)\dagger}({\bf k}) \, , \,  \hat{a}^{(\lambda^{\prime})\dagger}({\bf k}^{\prime}) \, \right]  = 0 
\; , \; 
\label{relcomua1}
\\
\left[ \, \hat{a}^{(\lambda)}({\bf k}) \, , \,  \hat{a}^{(\lambda^{\prime})\dagger}({\bf k}^{\prime}) \, \right]  &=& \delta^{\lambda\lambda^{\prime}}\, \delta^{3}({\bf k} - {\bf k}^{\prime}) \; .
\label{relcomua2}
\end{eqnarray}
\end{subequations}
In Coulomb gauge, the electric and magnetic field operators are $\hat{e}^{i}=-\partial_{t}\hat{a}^{i}$ and $\hat{b}^{i}=\epsilon^{ijk}\,\partial^{j}\hat{a}^{k}$, respectively.
The energy quantization for the linearized ED with external magnetic field is also promoting the conserved energy of the theory to an operator. 
We use the result of the conserved energy obtained in the ref. \cite{MJNevesPRD2021} whose the operator version is read as
\begin{eqnarray}\label{Hhat}
\hat{H}=\int d^{3}{\bf x} \, \frac{1}{2} \left[ \, \epsilon_{ij} \, \hat{e}_{i} \, \hat{e}_{j} +(\mu^{-1})_{ij} \, \hat{b}_{i} \, \hat{b}_{j} \, \right] \; ,
\end{eqnarray}
where $\epsilon_{ij}$ and $(\mu^{-1})_{ij}$ are the electric permittivity, and the inverse of magnetic permeability tensors
$\epsilon_{ij} = \delta_{ij}+d_{E}\,B_{i}\,B_{j}$ and $(\mu^{-1})_{ij} = \delta_{ij}-d_{B}\,B_{i}\,B_{j}$, respectively. The hamiltonian is positive 
if the coefficients satisfy the condition $d_{B}\,{\bf B}^2<1$. Substituting the expressions of $\hat{e}^{i}$ and $\hat{b}^{i}$ with the help of (\ref{solaiop}), we obtain the hamiltonian operator : 
\begin{eqnarray}\label{Hopintdk}
\hat{H} &=& -\int d^{3}{\bf k} \, \frac{1}{4} \, \omega({\bf k}) \left[\, 2\left(1-\frac{{\bf k}^2}{\omega^2} \right)+d_{E} \, (\hat{{\bf k}} \times {\bf B})^2+\frac{{\bf k}^2}{\omega^2} \, d_B \, (\hat{{\bf k}} \times {\bf B})^2 \, \right]
\nonumber \\
&&
\times\sum_{\lambda=1}^{2}\left[ \, \hat{a}^{(\lambda)}({\bf k})\, \hat{a}^{(\lambda)}(-{\bf k})\,e^{-i2\omega({\bf k})t}+\hat{a}^{(\lambda)\dagger}({\bf k})\, \hat{a}^{(\lambda)\dagger}(-{\bf k})\,e^{i2\omega({\bf k})t} \, \right] 
\nonumber \\
&&
+\int d^{3}{\bf k} \, \frac{1}{4} \, \omega({\bf k})\left[\, 2\left(1+\frac{{\bf k}^2}{\omega^2} \right)+d_{E} \, (\hat{{\bf k}} \times {\bf B})^2-\frac{{\bf k}^2}{\omega^2} \, d_B \, (\hat{{\bf k}} \times {\bf B})^2 \, \right] 
\nonumber \\
&&
\times\sum_{\lambda=1}^{2}\left[ \, \hat{a}^{(\lambda)}({\bf k})\, \hat{a}^{(\lambda)\dagger}({\bf k})+\hat{a}^{(\lambda)\dagger}({\bf k})\, \hat{a}^{(\lambda)}({\bf k}) \, \right] \; .
\end{eqnarray}
Defining the ground state as $\hat{a}^{(\lambda)}({\bf k})|0\rangle =0$, and the first excited state as $|1\rangle =\hat{a}^{(\lambda)\dagger}({\bf k})|0\rangle$, the lowest energy state is given by 
\begin{eqnarray}\label{E0}
\langle 0 |\hat{H}| 0 \rangle = \int d^{3}{\bf k} \, \frac{1}{4} \, \omega({\bf k})\left[\, 2\left(1+\frac{{\bf k}^2}{\omega^2} \right)+d_{E} \, (\hat{{\bf k}} \times {\bf B})^2-\frac{{\bf k}^2}{\omega^2} \, d_B \, (\hat{{\bf k}} \times {\bf B})^2 \, \right] \; ,
\end{eqnarray}
that we expect, the result is divergent. Notice that, there is not any approximation in the results (\ref{Hopintdk}) and (\ref{E0}) in relation to 
$\omega({\bf k})$-frequency. The approximation is just in the hamiltonian operator (\ref{Hhat}) in which we have the quadratic dependency 
in ${\bf e}$ and ${\bf b}$ due to linearized ED from section \ref{sec2}. The new physics here is dependency on the external magnetic field that affects the point zero energy of the linearized EM field with the direction of the wave propagation. In the situation of $\hat{{\bf k}}$ parallel to ${\bf B}$, the result (\ref{E0}) is reduced to usual ground state energy for the EM field.   
Other important point for discussion is that the hamiltonian operator depends on the time coordinate in the second line from (\ref{Hopintdk}). If we have a NLED in which 
$d_{E} \neq 0$, or $d_{B} \neq 0$, the terms of products $\hat{a}^{(\lambda)}({\bf k})\,\hat{a}^{(\lambda)}(-{\bf k})$ and $\hat{a}^{(\lambda)\dagger}({\bf k})\, \hat{a}^{(\lambda)\dagger}(-{\bf k})$ contributes to the hamiltonian operator. Thus, in general, we have $d\hat{H}/dt \neq 0$. The usual result is recovered when $d_{E} \rightarrow 0$ and $d_{B} \rightarrow 0$. The simplest case is when the $\omega$-frequency satisfies $\omega_{1}({\bf k})=|{\bf k}|$ in which the ground state energy is
\begin{eqnarray}\label{E0omega1}
\left. \langle 0 |\hat{H}| 0 \rangle \right|_{\omega_{1}=|{\bf k}|} = \int d^{3}{\bf k} \, |{\bf k}| \left[\, 1 + \frac{1}{4} \left(d_{E}-d_B\right) (\hat{{\bf k}} \times {\bf B})^2 \, \right] \; ,
\end{eqnarray}
in which if $d_{E}=d_{B}$ in some NLED, the effect of the external magnetic field disappears.

For end of this section, the microcausality is an fundamental property of a QFT defined by the commutator between two field-operators in different times.
It is related to the Pauli-Jordan (PJ) function that must be null for space-like intervals in the space-time. In the case of the linearized ED, the PJ 
function is defined by the commutator :
\begin{eqnarray}
i\,\Delta^{(PJ)}_{ij}(x-x^{\prime})=\left[ \, \hat{a}_{i}(x) \, , \, \hat{a}_{j}(x^{\prime}) \, \right] \; .
\end{eqnarray}
Using the Fourier representation (\ref{solaiop}) and the commutation relations (\ref{relcomua1})-(\ref{relcomua2}), the Pauli-Jordan function 
is set by  
\begin{eqnarray}\label{DeltaPJij}
\Delta^{(PJ)}_{ij}(x-x^{\prime})=-\,i\,\hat{P}_{ij} \, \Delta_{PJ}(x-x^{\prime}) \; ,
\end{eqnarray}
where 
\begin{eqnarray}\label{DeltaPJ}
\Delta_{PJ}(x-x^{\prime})=\int \frac{d^{3}{\bf k}}{(2\pi)^{3} \, 2 \, \omega({\bf k})} \, \left[ \, e^{-\,i\,k\cdot(x-x^{\prime})}-e^{\,i\,k\cdot(x-x^{\prime})} \, \right] \; .
\end{eqnarray}
Here, we have the scalar product $k\cdot(x-x^{\prime})=\omega({\bf k})(t-t^{\prime})-{\bf k}\cdot({\bf x}-{\bf x}^{\prime})$, in which $\omega({\bf k})$ satisfies
the dispersion relations (\ref{omegask}). Therefore, we have three possible PJ functions that must be null for the condition $(x-x^{\prime})^2 < 0$. As we expect, when $\omega_{1}({\bf k})=|{\bf k}|$ the PJ function satisfies the microcausality condition, but when the dispersion relations are given by $\omega_{2}({\bf k})$ and 
$\omega_{3}({\bf k})$ from (\ref{omegask}), this condition must be investigated. The calculus is not simple for any external magnetic field. We will show in the section (\ref{sec7}) that for the ModMax ED with an external magnetic field, the PJ function contains the microcausality condition. 
\section{Functional quantization of NLED under an external magnetic field}
\label{sec4}
The path integral quantization for the gauge theory of (\ref{L2simp}) is set by the generating functional
\begin{equation}\label{ZJ}
Z[J^{\mu}]=\!\!\int {\cal D}a^{\mu} \exp\left\{ i \int d^4x \left[ -\frac{1}{4} f_{\mu\nu}^{\, 2}
+\frac{d_B}{8} (F_{B\mu\nu}f^{\mu\nu})^{2}
+\frac{d_E}{8} (\widetilde{F}_{B\mu\nu}f^{\mu\nu})^{2}-\frac{1}{2\xi} (\partial_{\mu}a^{\mu})^{2}+J_{\mu}\,a^{\mu}  \right] \right\} \; ,
\end{equation}
where we have introduced the gauge fixing term with the $\xi$-real parameter. We fix the radiation gauge such that the time-like component of the gauge field 
is null, {\it i. e.}, $a^0=0$, that is equivalent to insert the Dirac delta $\delta[a^{0}]$ in the functional integration in (\ref{ZJ}). 
Thus, after the functional integration in $a^0$, the action inside the generating functional is written into form of field-operator-field
\begin{equation}\label{ZJi}
Z[J^{i}]=\!\!\int {\cal D}a^{i} \exp\left\{ i \int d^4x \left[ \, \frac{1}{2} \, a_{i} \, \hat{{\cal O}}_{ij}(\partial_{t},\partial_{i}) \, a_{j}+J_{i}\,a^{i} \, \right] \right\} \; ,
\end{equation}
%
%\begin{eqnarray}\label{L2aOa}
%{\cal L}^{(2)}=\frac{1}{2} \, a_{i} \, \hat{{\cal O}}_{ij}(\partial_{t},\partial_{i}) \, a_{j}-J_{i}\,a^{i} \; ,
%\end{eqnarray} 
%
in which the dynamics is transferred to the spatial component $a^{i}(x)\,(i=1,2,3)$, the functional integration is now performed over the transversal degree of freedom,
and the operator $\hat{{\cal O}}_{ij}$ is defined by
%and defined the operators
%
%\begin{eqnarray}
%&&
%\hat{{\cal O}}_{ij}(\partial_{t},\partial_{i})=\delta_{ij}\,\Box+\left( \frac{1}{\xi}-1 \right)\partial_{i} \, \partial_{j}
%\nonumber \\
%&&
%-d_B \left[ \, ({\bf B}\times{\bf \nabla})^2 \, \delta_{ij}-{\bf B}^2 \, \partial_{i}\,\partial_{j}- B_{i} \, B_{j} \, \nabla^2
%+({\bf B}\cdot{\bf \nabla})\left(B_{i}\,\partial_{j}+B_{j}\,\partial_{i}\right) \, \right]+d_{E} \, B_{i}\,B_{j}\, \partial_{t}^{\,2} 
%\nonumber \\
%&&
%-d_E \left[ \, ({\bf E}\times{\bf \nabla})^2 \, \delta_{ij}-{\bf E}^2 \, \partial_{i}\,\partial_{j}- E_{i} \, E_{j} \, \nabla^2
%+({\bf E}\cdot{\bf \nabla})\left(E_{i}\,\partial_{j}+E_{j}\,\partial_{i}\right) \, \right]+d_{B} \, E_{i}\,E_{j}\, \partial_{t}^{\,2} 
%\nonumber \\
%&&
%-2\,d_{B}\,E_{i} \, ({\bf B}\times\nabla)_{j}\,\partial_{t}-2\,d_{E}\,B_{i} \, ({\bf E}\times\nabla)_{j}\,\partial_{t}
%\; .
%\end{eqnarray}
%
%
\begin{eqnarray}\label{Oijspace}
\hat{{\cal O}}_{ij}(\partial_{t},\partial_{i}) &=& -\delta_{ij}\,\Box
+\left( \frac{1}{\xi}-1 \right)\partial_{i} \, \partial_{j}
-d_B \left[ \, ({\bf B}\times{\bf \nabla})^2 \, \delta_{ij}
-{\bf B}^2 \, \partial_{i}\,\partial_{j}
\right.
\nonumber \\
&&
\left.
\!\!\!
- B_{i} \, B_{j} \, \nabla^2
+({\bf B}\cdot{\bf \nabla})\left(B_{i}\,\partial_{j}+B_{j}\,\partial_{i}\right) 
\, \right]
+d_{E} \, B_{i}\,B_{j}\, \partial_{t}^{\,2} 
%\nonumber \\
%&&
%-d_E \left[ \, ({\bf E}\times{\bf \nabla})^2 \, \delta_{ij}
%-{\bf E}^2 \, \partial_{i}\,\partial_{j}
%- E_{i} \, E_{j} \, \nabla^2
%+({\bf E}\cdot{\bf \nabla})\left(E_{i}\,\partial_{j}+E_{j}\,\partial_{i}\right) 
%\, \right]+d_{B} \, E_{i}\,E_{j}\, \partial_{t}^{\,2} 
%-2\,d_{B}\,E_{i} \, ({\bf B}\times\nabla)_{j}\,\partial_{t}
%\nonumber \\
%&&
%-2\,d_{E}\,B_{i} \, ({\bf E}\times\nabla)_{j}\,\partial_{t}
\; .
\end{eqnarray}
%
%where we have used $\partial_{i}a^{i}=0$ inside the correspondent action when it is integrated by parts.
%
Using the solution (\ref{solamu}) for the radiation gauge, {\it i.e.},
\begin{eqnarray}\label{Solai}
a_{i}(x)=a_{0i}(x)+\int d^{4}x \, \Delta_{ij}(x-x^{\prime}) \, J_{j}(x^{\prime}) \; ,
\end{eqnarray}
the Green function $\Delta_{ij}(x-x^{\prime})$ is represented by the Fourier integral
\begin{eqnarray}\label{Deltaij}
\Delta_{ij}(x-x^{\prime})=\int \frac{d^4k}{(2\pi)^{4}} \, {\cal O}_{ij}^{\,-1}(k_0,{\bf k}) \, e^{-i\,k \cdot (x-x^{\prime})} \; .
\end{eqnarray}
where ${\cal O}_{ij}^{\,-1}(k_0,{\bf k})$ is the inverse of the operator (\ref{Oijspace}) in the momentum space. The usual representation 
of the derivatives operators in the momentum space are $k^{i} \rightarrow -i\,\partial^{i}$ and $k^{0} \rightarrow i\,\partial_{t}$, in which 
the $\hat{{\cal O}}_{ij}$-operator has the matrix elements in the $(k^{0},{\bf k})$-space : 
%
%with the integrating is given by 
%$\tilde{a}_{i} (k) \, {\cal O}_{ij}(k_{0},{\bf k}) \, \tilde{a}_{j}(k)/2$, where $\tilde{a}_{i}(k)$ is
%the Fourier transform of $a_{i}(x)$, and ${\cal O}_{ij}(k_0,{\bf k})$ are the matrix elements
%
\begin{eqnarray}\label{Oij}
{\cal O}_{ij} (k_{0},{\bf k}) =  \left[ \, k^2 +d_{B}\,({\bf B}\times{\bf k})^2
%+d_{E}\,({\bf E}\cdot{\bf k})^2 
\, \right] \delta_{ij}
%-\left(k_{0}^2-\frac{1}{\xi}\,{\bf k}^2-d_{B}\,{\bf k}^2\,{\bf B}^2-d_{E}\,{\bf k}^2\,{\bf E}^2\right) \omega_{ij} 
+\left( 1-\frac{1}{\xi}- d_{B} \, {\bf B}^2 \right) {\bf k}^2 \, \omega_{ij}
%\nonumber \\
%&&
%\hspace{-0.5cm}
%+ d_B \, ({\bf B}\cdot{\bf k})^2 \, Q^{B}_{Tij}
\nonumber \\
+\left(-d_{B}\, {\bf k}^2-d_{E} \, k_{0}^{\,2}\right) \, {\bf B}^2 \, P^{B}_{Lij}
+d_{B}\,({\bf k} \cdot {\bf B})^{2} \left( S_{ij}+Q_{ij} \right)
 \; ,
%\nonumber \\
%&&
%\hspace{-0.5cm}
%+ d_E \, ({\bf E}\cdot{\bf k})^2 \, Q^{E}_{Tij}
%+d_{E} \, {\bf k}^2 \, {\bf E}^2 \, P^{E}_{Tij}-d_{B} \, k_{0}^{\,2} \, {\bf E}^2 \, P^{E}_{Lij} 
%\nonumber \\
%&&
%\hspace{-0.5cm}
%+2\,d_{B}\, k_{0} \, Q_{ij} +2\,d_{E}\, k_{0} \, R_{ij} \; ,
\end{eqnarray}
in which $k^2=k_0^2-{\bf k}^2$ is the squared $4$-momentum, and we have 
%the old spatial projectors
%
%\begin{eqnarray}
%\theta_{ij}=\delta_{ij}-\frac{k_{i}\,k_{j}}{{\bf k}^2} 
%\; \; , \; \;
%\omega_{ij}=\frac{k_{i}\,k_{j}}{{\bf k}^2} \; ,
%\end{eqnarray}
%
defined the projectors 
%
%\begin{subequations}
\begin{eqnarray}\label{projectors}
\omega_{ij} = \frac{k_{i}\,k_{j}}{{\bf k}^2}
\; \; , \; \;
P^{B}_{Lij}=\frac{B_{i}\,B_{j}}{{\bf B}^2}
\; \; , \; \;
S_{ij}=\frac{B_{i}\,k_{j}}{{\bf k}\cdot{\bf B}}
\; \; , \; \;
Q_{ij}=\frac{k_{i}\,B_{j}}{{\bf k}\cdot{\bf B}} \; .
%\; , \;
%P^{E}_{Tij} = \delta_{ij}-\frac{E_{i}\,E_{j}}{{\bf E}^2}
%\; , \;
%P^{E}_{Lij}=\frac{E_{i}\,E_{j}}{{\bf E}^2}
%\; , \;
%\\
%Q_{ij} &=& E_{i}\, \epsilon_{jlm}\,B_{l}\,k_{m} 
%\; , \;
%R_{ij} = B_{i} \, \epsilon_{jlm}\,E_{l}\,k_{m} \; .
%Q^{E}_{Tij}=\delta_{ij}-\frac{k_{i}\,E_{j}}{{\bf k}\cdot{\bf E}}
%\; , \;
%Q^{E}_{Lij}=\frac{k_{i}\,E_{j}}{{\bf k}\cdot{\bf E}} \; .
\end{eqnarray}
%
%
%
%\end{subequations}
%
%Since that is known, $\theta_{ij}$ and $\omega_{ij}$ satisfy the relations
%
%\begin{eqnarray}
%\theta_{ik} \, \theta_{kj} = \theta_{ij} 
%\; \; , \; \;
%\omega_{ik} \, \omega_{kj}= \omega_{ij} 
%\; \; , \; \;
%\theta_{ik} \, \omega_{kj} = 0 \; .
%\end{eqnarray}
%
All the relations involving these projectors are illustrated in the table \ref{table1}. 
%satisfy the relations
%
%\begin{subequations}
%\begin{eqnarray}
%\omega_{ik}\,\omega_{kj}=\omega_{ij} \; , \; P^{B}_{Lik}\,P^{B}_{Lkj}=P^{B}_{Lij}  \; , \; \omega_{ik}\,P^{B}_{Lkj}=
%P^{B}_{T}\,Q^{B}_{L}=Q^{B}_{L}-P^{B}_{L} \; , 
%\\
%P^{B}_{L}\,P^{B}_{L}=P^{B}_{L} \; , \; P^{B}_{L}\,Q^{B}_{T}=0 \; , \; P^{B}_{L}\,Q^{B}_{L}=P^{B}_{L} \; , \; Q^{B}_{T}\,P^{B}_{T}=P^{B}_{T} \; , \; 
%\end{eqnarray}
%\end{subequations}
%%
%and
%

%%%   Table 1   %%%%%%%%%%%%%%%%%%%%%%%%%%%%%%%%%%%%%%%%%%%%%%%
%
\begin{table}[tb]
\centering
\begin{tabular}{|c||c|c|c|c|}
\hline
%after \\: \hline or \cline{col1-col2} \cline{col3-col4} ...
 &  ~$ \omega_{kj} $~ & ~$ P^{B}_{Lkj} $~ & ~$S_{kj}$~ & ~$Q_{kj}$ \\
\hline
\hline
$\omega_{ik}$  &  $\omega_{ij}$  &  $\cos^2\theta_{B} \, Q_{ij}$  &  $\omega_{ij}$  &  $Q_{ij}$  \\
\hline
$P^{B}_{Lik}$  &  $\cos^2\theta_{B} \, S_{ij}$ &  $P^{B}_{Lij}$  & $S_{ij}$ & $P^{B}_{Lij}$ \\
\hline
$S_{ik}$  & $S_{ij}$  &  $P^{B}_{Lij}$  &  $S_{ij}$ & $\sec^2\theta_{B} \, P^{B}_{Lij}$ \\
\hline
$Q_{ik}$ &  $\omega_{ij}$  &  $Q_{ij}$  &  $\sec^2\theta_{B} \, \omega_{ij}$  & $Q_{ij}$ \\
\hline
\hline
\end{tabular}
\caption{
The projectors algebra resumed in this table. The column of the projectors multiplies the projectors in the line above.
We define the $\theta_{B}$-angle as $\cos\theta_{B}=\hat{{\bf k}}\cdot\hat{{\bf B}}$.
}
\label{table1}
\end{table}
If we write the matrix elements (\ref{Oij}) as the combination ${\cal O}=a\,\mathds{1}+b\,\omega+c\,P_{L}^{B}+d\,S+e\,Q$, where 
the real coefficients $a$, $b$, $c$, $d$ and $e$ are given by
\begin{eqnarray}
a &=& k^2 +d_{B}\,({\bf B}\times{\bf k})^2 \, , \, b=\left( 1-\frac{1}{\xi}- d_{B} \, {\bf B}^2 \right) {\bf k}^2 \, , \,
\nonumber \\ 
c &=& \left(-d_{B}\, {\bf k}^2-d_{E} \, k_{0}^{\,2}\right) \, {\bf B}^2 \, , \, d=e=d_{B}\,({\bf k} \cdot {\bf B})^{2} \, ,
\end{eqnarray}
the inverse matrix ${\cal O}^{-1}$ is so written as the combination of the projectors (\ref{projectors})
\begin{equation}
{\cal O}^{-1} = a_{1} \, \mathds{1} + a_{2} \, \omega +a_{3} \, P_{L}^{B} + a_{4} \, S + a_5 \, Q \; .
%+ a_{4} \, P_{T}^{E}+a_{5} \, P_{L}^{E}+a_{6} \, Q+a_{7} \, R+ a_{8} \, U^{E}+a_{9} \, U^{B}
%+a_{10} \, S+a_{11} \, T \; ,
\end{equation}
The inverse condition ${\cal O}\,{\cal O}^{-1}=\mathds{1}$, and relations from the table (\ref{table1}), yield as results the coefficients 
\begin{eqnarray}\label{a1a2}
a_1 &=& \frac{1}{a}
\; , \;
a_2 = \frac{de-b(a+c)\cos^2\theta_{B}}{a[\,-de-bc\cos^4\theta_{B}+(a^2+bc+de+a(b+c+d+e))\cos^2\theta_{B}\,]}
\; , \;
\nonumber \\
a_{3} &=& \frac{de-c(a+b)\cos^2\theta_{B}}{a[\,-de-bc\cos^4\theta_{B}+(a^2+bc+de+a(b+c+d+e))\cos^2\theta_{B} \,]} \; ,
\nonumber \\
a_{4} &=& \frac{bc\cos^4\theta_{B}-d(a+e)\cos^2\theta_{B}}{a[\,-de-bc\cos^4\theta_{B}+(a^2+bc+de+a(b+c+d+e))\cos^2\theta_{B} \,]} \; ,
\nonumber \\
a_{5} &=& \frac{bc\cos^4\theta_{B}-e(a+d)\cos^2\theta_{B}}{a[\,-de-bc\cos^4\theta_{B}+(a^2+bc+de+a(b+c+d+e))\cos^2\theta_{B}\,]} \; ,
\end{eqnarray}   
where $\cos\theta_{B}=\hat{{\bf k}} \cdot \hat{{\bf B}}$ is the angle that ${\bf k}$-wave vector does with ${\bf B}$-direction. 
%Therefore, the Green function in the coordinate space is set by the Fourier transform
%

%
Fixing a radiation Feynman gauge with $\xi=1$, the coefficients (\ref{a1a2}) are simplified as
\begin{eqnarray}\label{a1a2xi=1}
a_1 &=& \frac{1}{k^2 +d_{B}\,({\bf B}\times{\bf k})^2}
\; , \;
a_2 = \frac{d_{B}\,{\bf k}^2\,{\bf B}^2}{k^2\,[\,k^2+d_{B}({\bf B}\times{\bf k})^2\,]}
\; , \;
\nonumber \\
a_{3} &=& \frac{{\bf B}^2}{k^2}\left[ \frac{d_{B}\,{\bf k}^2}{ k^2+d_{B}({\bf B}\times{\bf k})^2 } + \frac{d_{E}\,k_{0}^2}{k^2-d_{E}\,{\bf B}^2k_{0}^{2}} \right] 
\; , \;
%\nonumber \\
a_{4} = a_{5} = \frac{- \, d_{B}({\bf k}\cdot{\bf B})^{2}}{k^2 \, [\, k^2+d_{B}({\bf B}\times{\bf k})^2 \,]} \; ,
\end{eqnarray}     
and the expression of the Green function in the momentum space is read as
\begin{eqnarray}\label{propfoton}
\left. {\cal O}_{ij}^{\,-1}(k_0,{\bf k}) \right|_{B} &=& \frac{\delta_{ij}}{k^2+d_B({\bf k}\times{\bf B})^2}
+\frac{ d_B \, {\bf B}^2 }{k^2+d_B({\bf k}\times{\bf B})^2} \, \frac{k_{i}\,k_{j}}{k^2}
\nonumber \\
\hspace{-2cm}
&&
+ \left[ \, \frac{d_{B}\,{\bf k}^2}{k^2+d_B\left({\bf k}\times{\bf B}\right)^2} + \frac{d_{E}\,k_{0}^2}{k^2-d_E\,{\bf B}^2\,k_{0}^{2} } \, \right] \frac{ B_{i}\,B_{j} }{k^2}
\nonumber \\
\hspace{-2cm}
&&
- \frac{ d_{B}\,({\bf k}\cdot{\bf B}) }{k^2} \frac{k_{i}\,B_{j}+k_{j}\,B_{i}}{ k^2+d_B\left({\bf k}\times{\bf B}\right)^2 } \; .
%\nonumber \\
%\hspace{-1.5cm}
%&&
%-\frac{d_1d_2}{c_1} \, \frac{k_{0}^{2}}{k^2} \, \frac{ {\bf B}^2}{c_1k^2+d_2\,{\bf B}^2\,k_{0}^{2}} \, \frac{({\bf k}\cdot{\bf B})}{c_1k^2+d_1\left({\bf k}\times{\bf B}\right)^2} \, k_{i}\,B_{j} \; .
\end{eqnarray}
The limit $d_{B}=d_{E} \rightarrow 0$ recovers the usual result for the Green function in the momentum space in the Feynman gauge.     
The Green function (\ref{propfoton}) has poles at $k^{2}=0$, $k^2-d_E\,{\bf B}^2\,k_{0}^{2}=0$ and $k^2+d_{B}\left({\bf k}\times{\bf B}\right)^2=0$, 
that leads to the following frequencies
\begin{eqnarray}\label{polesB}
\omega_{\pm}^{(1)}=\pm \, |{\bf k}|
\; , \;
\omega_{\pm}^{(2)}=\pm \, |{\bf k}| \, \sqrt{1-d_B \, (\hat{{\bf k}}\times{\bf B})^{2}} 
\; , \;
\omega_{\pm}^{(3)}=\frac{\pm \, |{\bf k}|}{\sqrt{1-d_E\,{\bf B}^2}} \; ,
\end{eqnarray}
respectively. These results confirm two dispersions relations of the non-linear EDs under a magnetic background field showed in (\ref{omegask}), and $\omega_{\pm}^{(3)}$ are the new frequencies that emerge from the poles of the Green function in momentum space. In (\ref{polesB}), $\omega_{\pm}^{(1)}$ is associated with the gauge invariance of linearized ED, $\omega_{\pm}^{(2)}$ is a transverse mode like in the eq. (\ref{omegask}), and $\omega_{\pm}^{(3)}$ breaks the transversality if $d_{E} \neq 0$. All the frequencies are reduced to $\omega_{\pm}=\pm\,|{\bf k}|$ in the limit $d_{E}=d_{B} \rightarrow 0$. Other important point is that the Green function in the momentum space goes to zero in the ultraviolet limit of $k \rightarrow \infty$. Substituting the result (\ref{propfoton}) in (\ref{Deltaij}), the Green function in the coordinate space can be written as
\begin{eqnarray}
\Delta_{ij}(x-x^{\prime})=\hat{P}_{ij}\,\Delta(x-x^{\prime})+d_{E}\,B_{i}\,B_{j}\,\frac{\partial_{t}^2}{\Box}\,D(x-x^{\prime}) \; ,
\end{eqnarray}
where $\hat{P}_{ij}$ is the operator
\begin{eqnarray}
\hat{P}_{ij}=\delta_{ij}+d_{B}\, {\bf B}^2 \,\frac{\partial_{i}\,\partial_{j}}{\Box}-d_B\,\frac{({\bf B}\cdot\nabla)}{\nabla^2}\left(B_{i}\,\partial_{j}+B_{j}\,\partial_{i}\right) \; ,
\end{eqnarray}
and the functions $\Delta(x-x^{\prime})$ and $D(x-x^{\prime})$ are defined by the Fourier integrals
\begin{subequations}
\begin{eqnarray}
\Delta(x-x^{\prime}) &=& \int \frac{d^4k}{(2\pi)^2} \, \frac{e^{-ik\cdot(x-x^{\prime})}}{k^2+d_{B}({\bf k}\times{\bf B})^2} \; ,
\label{Delta}
\\
D(x-x^{\prime}) &=& \int \frac{d^4k}{(2\pi)^2} \, \frac{e^{-ik\cdot(x-x^{\prime})}}{k^2-d_{E}\,{\bf B}^2 \, k_{0}^2} \; .
\label{D}
\end{eqnarray}
\end{subequations} 
The Feynman propagator in the momentum space is obtained in (\ref{Delta}) and (\ref{D}) introducing the prescription $k^2 \rightarrow k^2+i\,\epsilon$, with $\epsilon>0$, and consequently, the poles (\ref{polesB}) are deviated in the complex plane of $k^{0}$ versus $|{\bf k}|$ to satisfy the causal Green function. The retarded and advanced Green functions are obtained through the prescription $k^{0} \rightarrow k^{0} \mp i\,\epsilon$, in which $(-)$ means the retarded Green function, and $(+)$ symbols the advanced Green function. 
The case with a pure electric background field is so obtained from result (\ref{propfoton}) making the substitutions ${\bf B} \rightarrow {\bf E}$, $d_{B} \rightarrow d_E$, and $d_{E} \rightarrow d_{B}$. The result is :
\begin{eqnarray}\label{propfotonE}
\left. {\cal O}_{ij}^{\,-1}(k_0,{\bf k})\right|_{E} &=& \frac{\delta_{ij}}{k^2+d_E({\bf k}\times{\bf E})^2}
+\frac{ d_E \, {\bf E}^2 }{k^2+d_E({\bf k}\times{\bf E})^2} \, \frac{k_{i}\,k_{j}}{k^2}
\nonumber \\
\hspace{-2cm}
&&
+ \left[ \, \frac{d_{E}\,{\bf k}^2}{k^2+d_E\left({\bf k}\times{\bf E}\right)^2} + \frac{d_{B}\,k_{0}^2}{k^2-d_B\,{\bf E}^2\,k_{0}^{2} } \, \right] \frac{ E_{i}\,E_{j} }{k^2}
\nonumber \\
\hspace{-2cm}
&&
- \frac{ d_{E}\,({\bf k}\cdot{\bf E}) }{k^2} \frac{k_{i}\,E_{j}+k_{j}\,E_{i}}{ k^2+d_E\left({\bf k}\times{\bf E}\right)^2 } \; ,
%\nonumber \\
%\hspace{-1.5cm}
%&&
%-\frac{d_1d_2}{c_1} \, \frac{k_{0}^{2}}{k^2} \, \frac{ {\bf B}^2}{c_1k^2+d_2\,{\bf B}^2\,k_{0}^{2}} \, \frac{({\bf k}\cdot{\bf B})}{c_1k^2+d_1\left({\bf k}\times{\bf B}\right)^2} \, k_{i}\,B_{j} \; .
\end{eqnarray}
whose the poles are
\begin{eqnarray}\label{polesE}
\omega_{\pm}^{(1)}=\pm \, |{\bf k}|
\; , \;
\omega_{\pm}^{(2)}=\frac{\pm \, |{\bf k}|}{\sqrt{1-d_B\,{\bf E}^2}}
\; , \;
\omega_{\pm}^{(3)}=\pm \, |{\bf k}| \, \sqrt{1-d_E \, (\hat{{\bf k}}\times{\bf E})^{2}} \; .
\end{eqnarray}

To conclude this section, if we substitute the solution (\ref{Solai}) in the expression (\ref{ZJi}), we obtain
\begin{equation}\label{ZJiJj}
Z[J_{i}]= \exp\left[ \, -\frac{i}{2} \int d^4x \, d^{4}x^{\prime} \, J_{i}(x)\,\Delta_{ij}(x-x^{\prime})\,J_{j}(x^{\prime}) \, \right] \; ,
\end{equation}
that has the standard form of the generating functional for the quantization of a free Abelian gauge theory in Coulomb gauge. Thus, we can obtain 
the correlation functions of $n$-points for a NLED in the presence of an external magnetic field, or of an external electric field.
\section{Effective potential in linearized NLEDs}
\label{sec5}
The path integral quantization of NLEDs in the presence of external EM fields opens the perspective to investigate the perturbation theory for the coupling with a complex 
scalar field. The scalar non-linear electrodynamics is set by the lagrangian :
\begin{equation}\label{Lscalar}
{\cal L}_{sc-nl}=(D_{\mu}\Phi)^{\ast} D^{\mu}\Phi-\mu^2 \, \Phi^{\ast}\Phi-\frac{\lambda}{6}\, (\Phi^{\ast}\Phi)^2
-\frac{1}{4} \, f_{\mu\nu}^{\, 2}
+\frac{d_B}{8} \, (F_{B\mu\nu}f^{\mu\nu})^{2}
+\frac{d_E}{8} \, (\widetilde{F}_{B\mu\nu}f^{\mu\nu})^{2} \; ,
\end{equation}
where $D_{\mu}=\partial_{\mu}+ig\, a_{\mu}$ is the covariant derivative operator that couples the complex scalar field $\Phi$ with the propagating 
$a^{\mu}$-potential of an NLED showed in the section \ref{sec2}, $g$ is the coupling constant, $\mu$ and $\lambda$ are two real parameters. 
The lagrangian (\ref{Lscalar}) is $U(1)$ gauge invariant under the local transformations : $\Phi \rightarrow \Phi^{\prime}(x)=e^{i\,\Lambda(x)}\,\Phi(x)$ and $a^{\mu} \rightarrow a^{\prime\mu}=a^{\mu}+\partial^{\mu}\Lambda$, for a real function $\Lambda$. We can define the generating functional for the connected correlation 
functions as usual : $W[J]=-i\,\ln Z[J]$. Thus, the correspondent effective action is given by the Legendre transformation $\Gamma[\phi_{c}]=W[J]-\int d^4x \, J(x) \, \phi_{c}(x)$, where the {\it classical field} is $\phi_{c}(x)=\delta W[J]/\delta J(x)$. Thereby, the effective potential formalism can be implemented in the model (\ref{Lscalar}). The scalar potential in (\ref{Lscalar}) has a minimal at $v=|\Phi_{0}|^2=\sqrt{- 6\,\mu^2/\lambda }$, for $\mu^2 < 0$, and we investigate the 
fluctuations of the complex scalar field around this minimum for the classical field $\phi_{c} \equiv v$.  
%
%Firstly, we use the previous expansion of gauge potential around the EM background, $A_{0\mu}(x)=a_{\mu}(x)+A_{B\mu}(x)$, with $A_{B\mu}(x)=-F_{B\mu\nu}\,x^{\nu}/2$, 
%in which the scalar model (\ref{Lscalar}) in this EM background is 
%
%As usual, the scalar sector of (\ref{Lscalar}) is
%
%\begin{equation}\label{Lsc2}
%{\cal L}_{sc}^{(2)}=\partial_{\mu}\phi^{\ast}\partial^{\mu}\phi+ie\left(\phi\,\partial_{\mu}\phi^{\ast}-\phi^{\ast}\partial_{\mu}\phi \right)(a^{\mu}+A_{B}^{\;\,\mu}) 
%+e^2\phi^{\ast}\phi\left(a_{\mu}+A_{B\mu} \right)^2
%-\mu^2 \phi^{\ast}\phi-\lambda\, (\phi^{\ast}\phi)^2 \; .
%\end{equation}
%
%
%The classical scalar field is defined as the vacuum expectation value of the scalar operator : $\phi_{c}=\langle 0 | \, \hat{\Phi} \, | 0 \rangle$. 
%
The complex scalar field is expanded as 
\begin{eqnarray}\label{phi}
\Phi(x)=\frac{1}{\sqrt{2}} \left[ \, \phi_c+\phi_1(x)+i\,\phi_{2}(x) \, \right] \; , 
\end{eqnarray}
for two real scalar fields $\phi_{a} \, (a=1,2)$. Substituting (\ref{phi}) in (\ref{Lscalar}), we obtain
\begin{eqnarray}\label{Lsc2phi}
{\cal L}_{sc} &=& \frac{1}{2}\, (\partial_{\mu}\phi_{1})^2-\frac{1}{2} \left( \, \mu^2+  \frac{\lambda}{2} \, \phi_{c}^2 \, \right) \phi_{1}^2+\frac{1}{2} \,  (\partial_{\mu}\phi_2)^2-\frac{1}{2} \left( \, \mu^2 + \frac{\lambda}{6} \, \phi_{c}^2\, \right) \phi_{2}^2
\nonumber \\
&&
+g\,\phi_{c} \, (\partial_{\mu}\phi_{2}) \,a^{\mu}  
+\frac{g^2}{2} \, \phi_{c}^2 \, a_{\mu}a^{\mu}
-V(\phi_{c})
+g\left(\phi_{1} \, \partial_{\mu} \phi_{2}-\phi_{2} \, \partial_{\mu}\phi_{1} \right) a^{\mu} 
\nonumber \\
&&
+g^2 \, \phi_{c} \, \phi_1\, a_{\mu}a^{\mu}
+ \frac{g^2}{2} \left( \, \phi_{1}^2 + \phi_{2}^{2} \, \right) a_{\mu}a^{\mu} 
-\left( \, \mu^2+\frac{\lambda}{6} \, \phi_{c}^2 \, \right)\phi_{c} \, \phi_{1}
\nonumber \\
&&
-\frac{\lambda}{6} \, \phi_{c} \, \phi_1 \left( \, \phi_{1}^2+\phi_{2}^{2} \, \right)-\frac{\lambda}{24} \left( \, \phi_{1}^2+\phi_{2}^2 \, \right)^2
 \; ,
\end{eqnarray}
where $V(\phi_c)=\mu^2 \, \phi_c^2/2+\lambda\,\phi_{c}^{4}/24$ is the effective potential at tree level for $\phi_{c}$. The gauge sector is so 
written in the form 
\begin{eqnarray}\label{Ln2quadratic}
{\cal L}_{nl}^{(2)}=\frac{1}{2} \, a_{\mu} \left[ \, \eta^{\mu\nu}\,\Box+ \left( \frac{1}{\xi}-1 \right)\partial^{\mu}\,\partial^{\nu}-d_{B}\, F_{B}^{\,\,\mu\alpha}\,F_{B}^{\,\,\nu\beta}\,\partial_{\alpha}\,\partial_{\beta} 
- d_{E}\, \widetilde{F}_{B}^{\,\,\mu\alpha}\,\widetilde{F}_{B}^{\,\,\nu\beta}\,\partial_{\alpha}\,\partial_{\beta}\, \right] a_{\nu} \, .
\end{eqnarray}
The first term in the second line from (\ref{Lsc2phi}) can be eliminated by integration inside the action. In the formulation of the perturbative QFT, only the quadratic terms contribute to the effective potential at one loop. The quadratic terms of (\ref{Lsc2phi}) and of (\ref{Ln2quadratic}) are joined, and fixing the Coulomb gauge with $\xi=1$, 
the action is written as 
\begin{equation}
S_{sc-nl}^{(2)} \left[\,\phi_{a} \, , \, a^{i} \, \right]=\int d^4x \, d^4x^{\prime} \left[ \, -\,\frac{1}{2} \, \phi_{a} (x) \, \hat{{\cal A}}_{ab}(x,x^{\prime};\phi_{c}) \, \phi_{b}(x^{\prime}) + \frac{1}{2} \, a_{i}(x) \, \hat{{\cal K}}^{ij}(x,x^{\prime};\phi_{c}) \, a_{j}(x^{\prime}) \right] \; ,
\end{equation}
where we have defined the operators :
\begin{subequations}
\begin{eqnarray}
\hat{{\cal A}}_{ab}(x,x^{\prime};\phi_{c}) &=& \left[ \, -\partial_{x\mu}\,\partial_{x^{\prime}}^{\mu} + m_{\phi_{a}}^2(\phi_{c}) \, \right] \delta_{ab} \, \delta^{4} (x-x^{\prime}) \; ,
\\
\hat{{\cal K}}^{ij}(x,x^{\prime};\phi_{c}) &=& \left[ \, - \, \delta^{ij} \left( \, -\partial_{x\alpha}\,\partial_{x^{\prime}}^{\alpha} + g^2 \, \phi_{c}^2 \, \right)
+d_{B}\, F_{B}^{\,\,i\alpha}\,F_{B}^{\,\,j\beta}\,\partial_{x\alpha}\,\partial_{x^{\prime}\beta} 
\right.
\nonumber \\
&&
\left.
+ d_{E}\, \widetilde{F}_{B}^{\,\,i\alpha}\,\widetilde{F}_{B}^{\,\,j\beta}\,\partial_{x\alpha}\,\partial_{x^{\prime}\beta} \, \right] \delta^{4} (x-x^{\prime}) \;  ,
\end{eqnarray}
\end{subequations}
with $m_{\phi_{1}}^2(\phi_{c}) = \mu^2+(\lambda/2)\,\phi_{c}^2$, and $m_{\phi_{2}}^2(\phi_{c}) =\mu^2+(\lambda/6)\,\phi_{c}^2$.  
 From the functional approach, the effective action at one loop is $\Gamma[\phi_c]=\Gamma_{0}[\phi_c]+\hbar \, \Gamma_{1}[\phi_c]$, where $\Gamma_{0}[\phi_c]=-\Omega\,V(\phi_{c})$ is the effective action at tree level, 
 and $\Gamma_{1}[\phi_c]$ is the correction at one loop
\begin{eqnarray}
\Gamma_{1}[\phi_c] = - \, \Omega \, V_{1}(\phi_c)=-\frac{i}{2} \left\{ \, \mbox{Tr} \ln\left[ \frac{ {\cal A}_{ab}(x,x^{\prime};\phi_c) }{{\cal A}_{ab}(x,x^{\prime};0)} \right] 
+\mbox{Tr} \ln\left[ \frac{ \hat{{\cal K}}^{ij}(x,x^{\prime};\phi_c) }{ \hat{{\cal K}}^{ij}(x,x^{\prime};0)} \right] \, \right\} \; ,
\end{eqnarray}
in which $\Omega$ is the volume of the space-time, 
$\mbox{Tr}$ denotes the trace over the operators in the coordinate space, and in the gauge sector, it is also included 
the trace over the spatial index $( i \, , \, j)$. 
In the momentum space, the effective potential at one loop is given by the integrals
\begin{eqnarray}\label{V1int}
V_{1}(\phi_c) &=& -\frac{i}{2}\int\frac{d^4k }{(2\pi)^{4}} \, \ln\left[\, 1-\frac{(\lambda/6)\,\phi_c^2}{k^2-\mu^2} \, \right]
-\frac{i}{2}\int\frac{d^4k }{(2\pi)^{4}} \, \ln\left[ \, 1-\frac{(\lambda/2)\,\phi_c^2}{k^2-\mu^2} \, \right]
\nonumber \\
&&
\hspace{-0.8cm}
-\frac{i}{2}\int\frac{d^4k }{(2\pi)^{4}} \, \ln\left[ \, 1-\frac{g^{2} \, \phi_c^2}{k^2} \, \right]
\nonumber \\
&&
\hspace{-0.8cm}
-\frac{i}{2}\int\frac{d^4k }{(2\pi)^{4}} \, \ln\left[ \frac{ (k^2-g^2 \, \phi_c^2)^2+(k^2-g^2 \,\phi_{c}^2)({\bf u}^2+{\bf w}^2)+{\bf u}^2\,{\bf w}^2 }{ (k^2)^2+k^2 \, ({\bf u}^2+{\bf w}^2) + {\bf u}^2 \, {\bf w}^2 } \, \right] \, , \; \;
\end{eqnarray}
where we have defined the vectors $u_{i}=\sqrt{d_{B}} \, F_{Bi\alpha}\,k^{\alpha}$ and $w_{i}=\sqrt{d_{E}} \, \widetilde{F}_{Bi\alpha}\,k^{\alpha}$. The integrals diverge in the ultraviolet limit, and 
%the factor $3$ in the second line means the three degree of freedoms for the massive vector gauge field. 
we use the dimensional regularization $(D)$ to make the previous integrals finite. Thus, we have 
\begin{eqnarray}\label{V1intD}
V_{1}(\phi_c,D) &=& -\frac{i}{2}\int\frac{d^Dk }{(2\pi)^{D}} \, \ln\left[\, 1-\frac{(\lambda/6) \, \Lambda^{2-D/2} \,\phi_c^2}{k^2-\mu^2} \, \right]
-\frac{i}{2}\int\frac{d^Dk }{(2\pi)^{D}} \, \ln\left[ \, 1-\frac{ (\lambda/2) \, \Lambda^{2-D/2} \,\phi_c^2}{k^2-\mu^2} \, \right]
\nonumber \\
&&
\hspace{-0.5cm}
-\frac{i}{2}\int\frac{d^Dk }{(2\pi)^{D}} \, \ln\left[ \, 1-\frac{g^{2} \Lambda^{2-D/2} \, \phi_c^2}{k^2} \, \right]
\nonumber \\
&&
\hspace{-0.8cm}
-\frac{i}{2}\int\frac{d^Dk }{(2\pi)^{D}} \, \ln\left[ \frac{ (k^2-g^2 \Lambda^{2-D/2} \, \phi_c^2)^2+(k^2-g^2 \Lambda^{2-D/2} \,\phi_{c}^2)({\bf u}^2+{\bf w}^2)+{\bf u}^2\,{\bf w}^2 }{ (k^2)^2+k^2 \, ({\bf u}^2+{\bf w}^2) + {\bf u}^2 \, {\bf w}^2 } \, \right] \; , \hspace{0.5cm}
\end{eqnarray}
where we have introduced the energy scale parameter $(\Lambda)$ in which the coupling constants remains dimensionless in $D$-dimensions. The integrals in the first line are calculated using the technics of the integrations from QFT, while in the second line we consider the background only with magnetic field on the 
${\cal Z}$-direction, such that, ${\bf B}=B\,\hat{{\bf z}}$. The spherical coordinates in $D$-dimensions leads the contractions of $k^{\mu}$ 
with the background tensor : ${\bf u}^2=({\bf B}\times{\bf k})^2=k^2\,B^2\,\sin^2\theta$ 
%$F_{Bi\alpha}\,F_{B}^{\,\,\,i\beta}\,k^{\alpha}\,k_{\beta}=({\bf B}\times{\bf k})^2=k^2\,B^2\,\sin^2\theta$ 
and ${\bf w}^2=-({\bf B}\cdot{\bf k})^2=-k^2\,B^2\,\cos^2\theta$,
%$\widetilde{F}_{Bi\alpha}\,\widetilde{F}_{B}^{\,\,\,i\beta}\,k^{\alpha}\,k_{\beta}=-({\bf B}\cdot{\bf k})^2=-k^2\,B^2\,\cos^2\theta$, 
with ${\bf k}$-vector on the radial direction. We show the calculus of the integrals in more details in the appendix. The result in $D$-dimensions is
\begin{eqnarray}\label{V1intDresult}
V_{1}(\phi_c,D) &=& \frac{-1}{2\,(4\pi)^{D/2}}\, \Gamma\left(-\frac{D}{2}\right) \left( 1+ 3^{D/2} \right) 
\left( \, \frac{\lambda}{6}\,\Lambda^{2-D/2}\,\phi_{c}^2 \, \right)^{D/2} 
\nonumber \\
&&
\hspace{-0.5cm}
-\frac{1}{2\,(4\pi)^{D/2}} \, \Gamma\left(-\frac{D}{2}\right) \, \left( \, g^2\,\Lambda^{2-D/2}\,\phi_{c}^2 \, \right)^{D/2} 
\nonumber \\
&&
\hspace{-0.5cm}
-\frac{1}{4\,(4\pi)^{D/2}} \, \Gamma\left(-\frac{D}{2}\right) \, \frac{\left( \, g^2\,\Lambda^{2-D/2}\,\phi_{c}^2 \, \right)^{D/2}}{\left[\,\left(1+d_{B}B^2\right)\left(1-d_{E}B^2\right)\,\right]^{(D-1)/2}} \; .
%\nonumber \\
\end{eqnarray}
%
%where ${}_2F_1$ is the Gauss hypergeometric function. 
The physics dimension is recovered writing $D=4-\epsilon$ with the expansion of the result (\ref{V1intDresult}) around $\epsilon \rightarrow 0$. We obtain   
\begin{eqnarray}\label{V1intDresultepsilon}
V_{1}(\phi_c,\epsilon) &=& -\frac{5 \, \lambda^2 \, \phi_c^{4}}{576\pi^2}\frac{1}{\epsilon} 
+ \frac{5\,\lambda^2 \, \phi_c^{4}}{2304 \pi^2} \left[\, -3+2\,\gamma_{E}+\frac{9}{5}\,\ln(3)+2 \,\ln\left( \frac{\lambda\,\phi_{c}^2}{4\pi\,\Lambda^2} \right) \, \right]
\nonumber \\
&&
\hspace{-0.5cm}
+\frac{g^4\,\phi_c^4}{32\pi^2} 
\left(-\frac{1}{\epsilon}+\frac{1}{4}\right)\left\{ \, -3+2\,\gamma_{E}+ 2\ln\left[ \, \frac{g^2 \, \phi_c^2}{4\pi \, \Lambda^2} \, \right] \, \right\}
\nonumber \\
&&
\hspace{-0.5cm}
+\frac{g^4\,\phi_c^4}{64\pi^2\left[\,\left(1+d_B B^2\right) \left(1-d_{E} B^2\right)\,\right]^{3/2}} 
\left(-\frac{1}{\epsilon}+\frac{1}{4}\right)
\nonumber \\
&&
\hspace{-0.5cm}
\times \left\{ \, -3+2\,\gamma_{E}+ 2\ln\left[ \, \frac{g^2 \, \phi_c^2}{4\pi \, \Lambda^2\left(1+d_B B^2\right) \left(1-d_E B^2\right)} \, \right] \, \right\}
 \; ,
\end{eqnarray} 
where the $\mu^2$-parameter is very small in relation to $\lambda\,\phi_{c}^2$ inside the loop integrals, and $\gamma_{E}=0.577$ is the Euler-Mascheroni constant. 
Summing this result to the potential at tree level, we obtain the effective potential with the correction at one loop :
\begin{eqnarray}\label{Veff1result}
V_{eff}^{(1)}(\phi_{c},\epsilon) &=& 
\frac{1}{2} \, \mu^2 \, \phi_c^2+
\frac{1}{24} \, \lambda\,\phi_{c}^{4}
-\frac{5 \, \lambda^2 \, \phi_c^{4}}{576\pi^2}\frac{1}{\epsilon} 
+ \frac{5\,\lambda^2 \, \phi_c^{4}}{2304 \pi^2} \left[\, -3+2\,\gamma_{E}+\frac{9}{5} \ln(3)+2 \,\ln\left( \frac{\lambda\,\phi_{c}^2}{4\pi\,\Lambda^2} \right) \, \right]
\nonumber \\
&&
\hspace{-0.5cm}
+\frac{g^4\,\phi_c^4}{32\pi^2} 
\left(-\frac{1}{\epsilon}+\frac{1}{4}\right)\left\{ \, -3+2\,\gamma_{E}+ 2\ln\left[ \, \frac{g^2 \, \phi_c^2}{4\pi \, \Lambda^2} \, \right] \, \right\}
\nonumber \\
&&
\hspace{-0.5cm}
+\frac{g^4\,\phi_c^4}{64\pi^2\left[\,\left(1+d_B B^2\right) \left(1-d_{E} B^2\right)\,\right]^{3/2}} 
\left(-\frac{1}{\epsilon}+\frac{1}{4}\right)
\nonumber \\
&&
\hspace{-0.5cm}
\times \left\{ \, -3+2\,\gamma_{E}+ 2\ln\left[ \, \frac{g^2 \, \phi_c^2}{4\pi \, \Lambda^2\left(1+d_B B^2\right) \left(1-d_E B^2\right)} \, \right] \, \right\} \; .
\end{eqnarray}
This result, as we expect, is divergent (non-physical result) by the presence of the term $1/\epsilon$, taking the limit $\epsilon \rightarrow 0$. Thereby, we need to renormalize it in terms of the physical parameters, and also introducing the counter-terms needed to remove the divergences. We will do it in the next section.  
\section{Renormalization of the effective potential}
\label{sec6}
Since we know from the previous section, the one loop integrals that yield effective potential at one loop are divergent in $D=4$, and we have used the 
dimensional regularization to extract the divergent terms and also the physicals terms that contribution to the effective potential. The result (\ref{Veff1result}) 
is divergent, and for the renormalization procedure, we need to redefine all the parameters of the model (fields and coupling constants) in terms of the physical 
parameters and also adding the counter-terms needed to remove the divergences. Thereby, we rewrite the one loop effective potential in terms of physical fields and 
constant couplings in the form  
\begin{eqnarray}\label{Veff1resultrenor}
V_{eff}^{(1)}(\phi_{c},\epsilon) &=& 
\frac{1}{2} \, \mu^2 \, \phi_c^2+
\frac{1}{24} \, \left(\lambda+\delta\lambda\right)\,\phi_{c}^{4}
-\frac{5 \, \lambda^2 \, \phi_c^{4}}{576\pi^2}\frac{1}{\epsilon} 
\nonumber \\
&&
\hspace{-0.5cm}
+ \frac{5\,\lambda^2 \, \phi_c^{4}}{2304 \pi^2} \left[\, -3+2\,\gamma_{E}+\frac{9}{5}\,\ln(3)+2 \,\ln\left( \frac{\lambda\,\phi_{c}^2}{4\pi\,\Lambda^2} \right) \, \right]
\nonumber \\
&&
\hspace{-0.5cm}
+\frac{g^4\,\phi_c^4}{32\pi^2} 
\left(-\frac{1}{\epsilon}+\frac{1}{4}\right)\left\{ \, -3+2\,\gamma_{E}+ 2\ln\left[ \, \frac{g^2 \, \phi_c^2}{4\pi \, \Lambda^2} \, \right] \, \right\}
\nonumber \\
&&
\hspace{-0.5cm}
+\frac{g^4\,\phi_c^4}{64\pi^2\left[\,\left(1+d_B B^2\right) \left(1-d_{E} B^2\right)\,\right]^{3/2}} 
\left(-\frac{1}{\epsilon}+\frac{1}{4}\right)
\nonumber \\
&&
\hspace{-0.5cm}
\times \left\{ \, -3+2\,\gamma_{E}+ 2\ln\left[ \, \frac{g^2 \, \phi_c^2}{4\pi \, \Lambda^2\left(1+d_B B^2\right) \left(1-d_E B^2\right)} \, \right] \, \right\}
 \; ,
\end{eqnarray}
where 
%$\delta \mu^2$ and 
$\delta\lambda$ is the counter-term that renormalizes the $\phi_{c}^4$-vertex. This counter-term is obtained through the renormalization condition :
\begin{eqnarray}
%\left. \frac{d^2V_{eff}}{d\phi_{c}^2} \right|_{\phi_c=0} = \mu^2
%\hspace{0.5cm} \mbox{and} \hspace{0.5cm}
\left. \frac{d^4V_{eff}}{d\phi_{c}^4} \right|_{\phi_c=M} = \lambda \; ,
\end{eqnarray}
for an arbitrary energy scale $M$, and $\lambda$ is the renormalized $\phi_{c}^4$-coupling. Using this condition in (\ref{Veff1resultrenor}), 
the $\delta\lambda$-counter-term is given by
\begin{eqnarray}
&&
\delta\lambda =
\frac{5 \, \lambda^2}{24 \pi^2}\frac{1}{\epsilon} 
-\frac{5 \, \lambda^2}{18 \pi^2}-\frac{5 \, \lambda^2 \, \gamma_{E}}{48 \pi^2}
-\frac{3 \, \lambda^2}{32 \pi^2}\,\log (3)-\frac{5 \, \lambda^2}{48 \pi^2}\, \ln \left(\frac{\lambda \,  M^2}{24 \pi  \Lambda^2}\right)
\nonumber \\
&&
+\frac{g^4}{\pi^2} \left( \frac{3}{4\,\epsilon}-1-\frac{3\,\gamma_{E}}{8} \right)
-\frac{3 \, g^4}{8 \pi^2}
\ln \left[\frac{g^2 \, M^2}{4 \pi \Lambda^2}\right] 
\nonumber \\
&&
+\frac{g^4}{2 \pi^2 \left[ \, \left(1+d_{B}B^2\right) \left(1-d_{E}B^2\right) \, \right]^{3/2}} \left( \frac{3}{4\,\epsilon}-1-\frac{3\,\gamma_{E}}{8} \right)
\nonumber \\
&&
-\frac{3\, g^4}{16 \pi^2 \left[ \, \left(1+d_{B}B^2\right) \left(1-d_{E}B^2\right) \, \right]^{3/2}}
\ln \left[\frac{g^2 \, M^2}{4 \pi \Lambda^2 \left(1+d_{B}B^2\right) \left(1-d_{E}B^2\right)}\right] \; .
\end{eqnarray}
Substituting this result in (\ref{Veff1resultrenor}), we obtain
\begin{eqnarray}\label{Veff1finite}
V_{eff}^{(1)}(\phi_c)&=&
\frac{1}{2}\,\mu^2\,\phi_c^{2}+
\frac{\lambda}{24} \, \phi_{c}^4+\frac{5\,\lambda^2 \, \phi_{c}^{4}}{1152\pi^2} \left[ \, \ln\left( \frac{\phi_c^2}{M^2} \right) -\frac{25}{6} \, \right]
\nonumber \\
&&
\hspace{-0.5cm}
+\frac{g^4\,\phi_c^4}{64\pi^2} \left\{ 1+\frac{1}{2\left[ \, \left(1+d_{B}B^2\right) \left(1-d_{E}B^2\right) \, \right]^{3/2} } \right\}
\left[ \, \ln \left( \frac{\phi_c^2}{M^2} \right)-\frac{25}{6} \, \right] \; .
\end{eqnarray}

It is worth to note that this result is valid for any NLED in the presence of an external magnetic field. All information about the NLED is provided in the
coefficients $d_E$ and $d_B$. When $\left( \, d_{B} \, , \, d_{E} \,\right) \rightarrow 0$ (or when $B \rightarrow 0$), the effective potential for the scalar QED 
is recovered in the Coulomb gauge, that is known as the Coleman-Weinberg effective potential. We apply this result in the ModMax ED in the next section.
\section{Application in ModMax electrodynamics}
\label{sec7}
The ModMax electrodynamics is set by the lagrangian
\begin{eqnarray}\label{ModMaxL}
{\cal L}_{MM}({\cal F}_{0},{\cal G}_{0})=\cosh\gamma \, {\cal F}_0 + \sinh\gamma \, \sqrt{{\cal F}_0^2+{\cal G}_0^2} \; ,
\end{eqnarray}
where $\gamma \geq 0$ is a real parameter that must be positive to insure the causality and the unitarity at tree level of the theory. 
The Maxwell ED is recovered in (\ref{ModMaxL}) taking the limit $\gamma \rightarrow 0$. We show now the results in which the medium governed by the 
ModMax ED is submitted to an external magnetic field.  
Using the expansion discussed in the section \ref{sec2}, the coefficients under an uniform magnetic background field are given by
\begin{eqnarray}
d_{B}=\frac{d_1}{c_1} = 0
\quad \mbox{and} \quad
d_{E} = \frac{d_2}{c_1} =  2\,\frac{e^{\gamma}\sinh\gamma}{{\bf B}^2} \; . 
\end{eqnarray}
The Green function (\ref{propfoton}) is reduced to expression
\begin{eqnarray}\label{propfotonModMaxB}
\left. {\cal O}_{ij}^{\,-1}(k_0,{\bf k})\right|_{B} = \frac{\delta_{ij}}{k^2}
+ \frac{e^{\gamma}-1}{\left(2-e^{2\gamma}\right) k_{0}^2-{\bf k}^2} \frac{k_{0}^{2}}{k^2} \frac{ B_{i}\,B_{j} }{{\bf B}^2} \; .
\end{eqnarray}
In the coordinate space is given by
\begin{eqnarray}
\Delta_{ij}(x-x^{\prime})=\delta_{ij} \, \Delta(x-x^{\prime})+ 2 e^{\gamma} \sinh\gamma \, \frac{B_{i}\,B_{j} }{ {\bf B}^2}  \,\frac{\partial_{t}^2}{\Box}\,D(x-x^{\prime}) \; ,
\end{eqnarray}
%
%where $\hat{P}_{ij}$ is the operator
%
%\begin{eqnarray}
%\hat{P}_{ij}=\delta_{ij}+d_{B}\, {\bf B}^2 \,\frac{\partial_{i}\,\partial_{j}}{\Box}-d_B\,\frac{({\bf B}\cdot\nabla)}{\nabla^2}\left(B_{i}\,\partial_{j}+B_{j}\,\partial_{i}\right) \; ,
%\end{eqnarray}
%
where $\Delta(x-x^{\prime})$ is the usual Green function for a massless scalar field, and $D(x-x^{\prime})$ is 
%
%\begin{subequations}
\begin{eqnarray}
%\Delta(x-x^{\prime}) &=& \int \frac{d^4k}{(2\pi)^2} \, \frac{e^{-ik\cdot(x-x^{\prime})}}{k^2+d_{B}({\bf k}\times{\bf B})^2} \; ,
D(x-x^{\prime}) = \int \frac{d^4k}{(2\pi)^2} \, \frac{e^{-ik\cdot(x-x^{\prime})}}{ \left(2-e^{2\gamma}\right) k_{0}^2-{\bf k}^2} \; .
\label{DMM}
\end{eqnarray}
%\end{subequations} 
% 
%
The retarded Green function is set by the poles deviated by $k_{0}^{\pm}=\pm\,\omega_{E}({\bf k})-i\,\epsilon$, with $\omega_{E}({\bf k})=|{\bf k}|/\sqrt{ 2-e^{2\gamma} }$. Using the appropriated contour in the complex plane of $k^{0}$ for $t-t^{\prime}>0$, the integrals of (\ref{DMM}) yields the result
\begin{eqnarray}\label{DretMM}
D^{(-)}(x-x^{\prime})=\frac{\sqrt{2-e^{2\gamma}}}{2\pi} \, \delta\left[ \, \frac{(t-t^{\prime})^2}{2-e^{2\gamma}}-|{\bf x}-{\bf x}^{\prime}|^2 \, \right] \,\Theta(t-t^{\prime}) \; .
\end{eqnarray}
The Green function is null for the condition
\begin{eqnarray}
\frac{(t-t^{\prime})^2}{2-e^{2\gamma}}-|{\bf x}-{\bf x}^{\prime}|^2 \neq 0 \; ,
\end{eqnarray}
in which the quadratic intervals of the coordinates are modified by the presence of the $\gamma$-parameter. This result insures that $D^{(-)}(x-x^{\prime})=0$ for intervals outside the light-cone, thus, it respects the causality. The retarded time for the propagation of a signal is given by $t_{r}=t-\sqrt{2-e^{2\gamma}}\,|{\bf x}-{\bf x}^{\prime}|$, that constraints the condition $0< \gamma < 0.34$. 
The results of the canonical quantization applied to the lagrangian  (\ref{ModMaxL}) yield the ground state energy
\begin{eqnarray}\label{E0MM}
\langle 0 |\hat{H}| 0 \rangle = \int d^{3}{\bf k} \, \frac{1}{4} \, \omega({\bf k})\left[\, 2\left(1+\frac{{\bf k}^2}{\omega^2} \right)+ \left(e^{2\gamma}-1\right) \sin^2\theta_{B} \, \right] \; ,
\end{eqnarray}
where the solutions for the dispersion relations are : $\omega_{1}({\bf k})=\omega_{2}({\bf k}) = |{\bf k}|$, and $\omega_{3}({\bf k})=|{\bf k}| \sqrt{ 1- \left(1-e^{-2\gamma}\right) \sin^2\theta_{B} }$. These results do not depend on the magnetic field, but it does with the $\theta_{B}$-angle, namely, $\cos\theta_{B}=\hat{{\bf k}} \cdot \hat{{\bf B}}$. Evaluating (\ref{E0MM}) in these frequencies, for $\gamma \ll 1$, we obtain
%
%\begin{subequations}
\begin{eqnarray}\label{E0MMomega13}
\left. \langle 0 |\hat{H}| 0 \rangle \right|_{\omega_{1}}=\left. \langle 0 |\hat{H}| 0 \rangle \right|_{\omega_{3}} 
= \int d^{3}{\bf k} \, |{\bf k}| \, \left(\, 1+ \frac{\gamma}{2} \, \sin^2\theta_{B} \, \right) \; .
%\\
%\left. \langle 0 |\hat{H}| 0 \rangle \right|_{\omega_{3}} &=&  \int d^{3}{\bf k} \, |{\bf k}| \, \left(\, 2+ 3\,\gamma\, \sin^2\theta_{B} \, \right)
\end{eqnarray}
%\end{subequations}
%

\begin{figure}[t]
\centering
\includegraphics[width=0.75\linewidth]{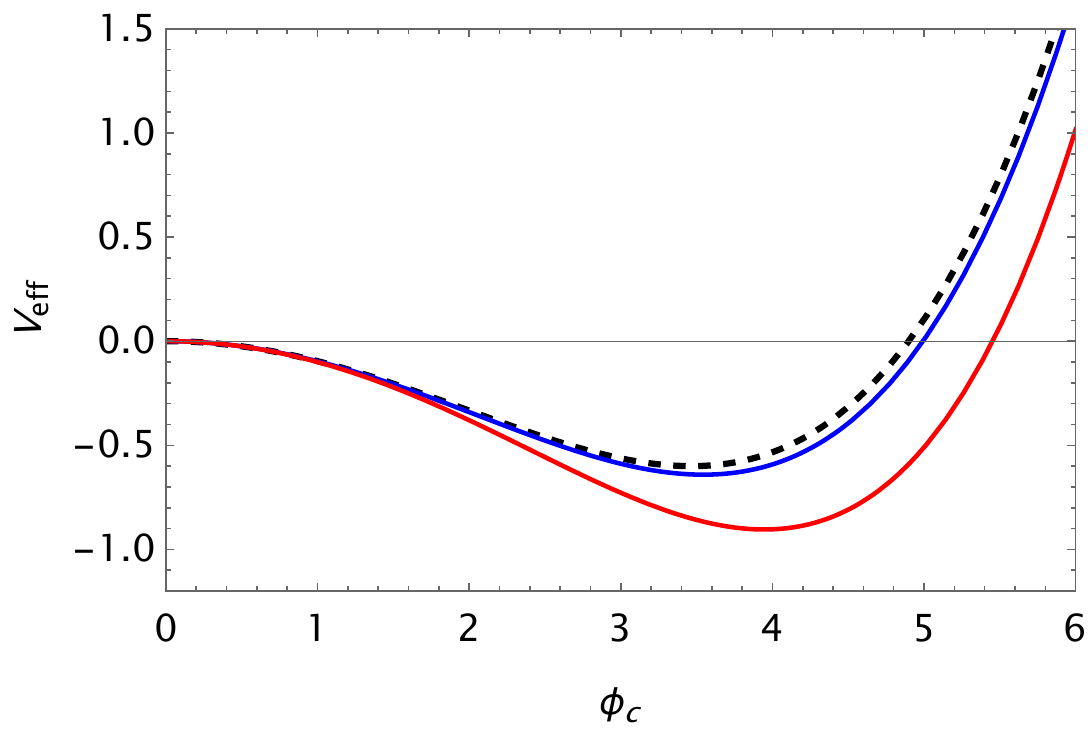}
\caption{The tree level (black dashed line) and the renormalized effective potential (one loop) as functions of $\phi_c$. 
In this plot, we use $\mu^2=-0.2$, $\lambda=0.1$, $M=1$ (energy dimension) and $g=0.5$, for the values of $\gamma=0.06$ (blue line) and $\gamma=0.31$ (red line).}
    \label{fig1}
\end{figure}

%
The microcausality analysis starts substituting the dispersion relation $\omega_{3}({\bf k})$ in the PJ function (\ref{DeltaPJ}), we obtain the integrals
\begin{eqnarray}\label{DeltaPJMM}
\Delta_{PJ}(x-x^{\prime})=\frac{1}{8\pi^2}\int_{0}^{\infty} dk \, k \int_{0}^{\pi} d\theta \, \frac{\sin\theta}{\sqrt{ \cos^2\theta +e^{-2\gamma} \sin^2\theta }} 
\nonumber \\ 
\times \left[ \, e^{-\,i\,k\,\tau \, \sqrt{ \cos^2\theta +e^{-2\gamma} \sin^2\theta}+i \, k \, R\cos\theta}-e^{i\,k\,\tau \, \sqrt{ \cos^2\theta +e^{-2\gamma} \sin^2\theta}-i \, k \, R\cos\theta} \, \right] \; ,
\end{eqnarray}
where $\tau=t-t^{\prime}$ and $R=|{\bf x}-{\bf x}^{\prime}|$ are real and positive intervals, and we are considering the magnetic field on ${\cal Z}$-direction. With $\gamma \ll 1$, the integrals yield the result
\begin{eqnarray}\label{DeltaPJMMresult}
&&
\Delta_{PJ}(x-x^{\prime})=-\frac{i}{4\pi R}\left[ \, \delta(R-\tau)-\delta(R+\tau) \, \right]
\nonumber \\
&&
-\frac{i\,\gamma}{2\pi R^3}\left\{ \, R\,\tau \left[ \, \delta(R-\tau)+\delta(R+\tau) \, \right]-\frac{R}{2}\left[ \, \mbox{sgn}(\tau+R)+\mbox{sgn}(\tau-R) \, \right]
\right.
\nonumber \\
&&
\left.
-\frac{\tau}{2}\left[ \, \mbox{sgn}(\tau+R)+\mbox{sgn}(R-\tau) \, \right]+\frac{1}{2} \left( \, R+\tau-|R-\tau| \, \right)
 \, \right\} \; .
\end{eqnarray}
For space-like intervals $(x-x^{\prime})^2 < 0$, that leads to $R > \tau$, the result (\ref{DeltaPJMMresult}) is null : 
\begin{eqnarray}\label{PJfunctionMM}
\Delta_{PJ}(x-x^{\prime})= 0-\frac{i\,\gamma}{2\pi R^3} \left( 0-0-\frac{\tau}{2} \, 2+\frac{1}{2} \, 2\tau \right) =0 \; .
\end{eqnarray}
Thereby, we show that the term with the contribution of the ModMax $\gamma$-parameter does not affect the microcausality condition.  
Now we apply the ModMax ED in the renormalized effective potential (\ref{Veff1finite}). The result is
\begin{eqnarray}\label{Veff1finiteModMax}
V_{eff}^{(1)}(\phi_c)&=&
\frac{1}{2}\,\mu^2\,\phi_c^{2}+
\frac{\lambda}{24} \, \phi_{c}^4+\frac{5\,\lambda^2 \, \phi_{c}^{4}}{1152\pi^2} \left[ \, \ln\left( \frac{\phi_c^2}{M^2} \right) -\frac{25}{6} \, \right]
\nonumber \\
&&
\hspace{-0.5cm}
+\frac{g^4\,\phi_c^4}{64\pi^2} \left[ \, 1+\frac{1}{2\left(2-e^{2\gamma}\right)^{3/2} } \, \right]
\left[ \, \ln \left( \frac{\phi_c^2}{M^2} \right)-\frac{25}{6} \, \right] \; .
\end{eqnarray}
%
%\begin{eqnarray}\label{Veff1finiteModMax}
%V_{eff}^{(1)}(\phi_c)&=&\frac{1}{2}\,\mu^2\,\phi_c^{2}+\frac{\lambda}{24} \, \phi_{c}^4+\frac{5\,\lambda^2 \, \phi_{c}^{4}}{1152\pi^2} \left[ \, \ln\left( \frac{\phi_c^2}{M^2} %\right) -\frac{25}{6} \, \right]
%\nonumber \\
%&&
%\hspace{-0.5cm}
%+\frac{3\,g^4\,\phi_c^4}{128\pi^2 \left[ \, 1-2\,e^{\gamma}\sinh(\gamma)/3 \, \right]^{3/2}} \left[ \, \ln \left( \frac{\phi_c^2}{M^2} \right)-\frac{25}{6} \, \right] \; ,
%\end{eqnarray}
%
that does not depend on the magnetic field, and imposes that $0< \gamma < 0.34$. The renormalized effective potential is plotted as function of $\phi_c$ in the figure (\ref{fig1}) in comparison with the tree level potential set by the black dashed line. The parameters used for this plot are : $\mu^2=-0.2$, $\lambda=0.1$,  $M=1$ (in energy dimension), and $g=0.5$. The curves illustrate the values of $\gamma=0.06$ (blue line) and $\gamma=0.31$ (red line) for the ModMax parameter. The figure shows the minimal of the effective potential decreasing with the values of the ModMax parameter. As consequence, if the ModMax's non-linearity is strong, it favours for a minimal of the effective potential decreasing the stable VEV.        
The PVLAS (Polarization Vacuum with Laser) experiment estimates the vacuum birefringence with refractive index of 
$\Delta n_{B}/B^2=(19\pm 27) \times 10^{-24} \, \mbox{T}^{-2}$, for an external magnetic field of $B=2.5$ T \cite{Ejlli}. Using this estimative 
in the ModMax ED, the $\gamma$-parameter has the upper bound $|\gamma| < 1.18 \, \times \, 10^{-22}$ \cite{nossoJHEP2}. Since this value is very small, 
the effects in the results (\ref{DretMM}), (\ref{E0MM}) and (\ref{Veff1finiteModMax}) are negligible. In the case of Pauli-Jordan function,
the PVLAS bound for $\gamma$-parameter is compatible with the result (\ref{PJfunctionMM}).

%in which divergent terms emerge when $\Lambda \rightarrow \infty$.

%******************

%the Coleman-Weinberg effective potential at one loop with the contribution of a non-linear ED through the $d_{E}$-coefficient, 
%and of the external magnetic field. When $d_{E} \rightarrow 0$ (or when $B \rightarrow 0$), the effective potential for the scalar QED 
%is recovered in the Coulomb gauge. The other particular case is when the magnetic field is perpendicular to ${\bf k}$-direction. 
%In this case $({\bf k}\times{\bf B})^2=k^2\,B^2$ and $({\bf k}\cdot{\bf B})^2=0$, where the result for the integrals is similar, 
%exchanging the $d_{E}$-coefficient by $(-\,d_{B})$ in (\ref{V1intD}). 

%\newpage

%
\section{Concluding Comments}
\label{sec8}
In this paper, we show the canonical and path integral quantization approaches for an non-linear electrodynamics (NLEDs) 
in the presence of an magnetic background field. From a general lagrangian density that is function of the Lorentz- and gauge-invariants 
${\cal F}=-F_{\mu\nu}^2/4$ and ${\cal G}=-F_{\mu\nu}\widetilde{F}^{\mu\nu}/4$, we expand the gauge potential around an electromagnetic 
background field, considered uniform and constant. For small fluctuations up to second order in propagating electromagnetic fields, we obtain 
the linearized lagrangian density that governs the dynamic of this electromagnetic field in terms of coefficients that depends on the 
background EM field and also depends on the parameter of the NLED. We have investigated the canonical quantization of any linearized ED in 
an external magnetic field. Promoting the propagating fields to operators, the rules of the canonical quantization are imposed in which 
the hamiltonian operator is an temporal function in the terms of $\hat{a}\,\hat{a}$ and $\hat{a}^{\dagger}\,\hat{a}^{\dagger}$, for the 
annihilation and creation operators. In the ground state, the fundamental energy depends on the angle that the magnetic background field does 
with the wave vector, and also on the dispersion relation of the general solution of the plane wave, as we expect. 
For the simplest frequency solution, the fundamental energy depends on $(\hat{{\bf k}}\times{\bf B})^2$, in which this term 
disappears in the limit of the linear ED, as it is showed in the result (\ref{E0omega1}). 
Posteriorly, the path integral quantization is proposed building the generating functional for the linearized ED in a magnetic background field. 
In the Coulomb gauge, we obtain the inversion of the quadratic operator in the propagating fields, where the causal Green function is achieved in terms 
of the external magnetic field. This result allows to obtain the generating functional in the standard form, and then it is possible construct all the 
perturbative formalism of QFT. As application of this formalism, we calculate the effective potential at one loop for an any 
linearized ED coupled to complex scalar field. Since the scalar potential develops a VEV, we expand the scalar sector around the classical field, and the 
functional integrations for the effective action yield the one loop contribution to the effective potential. The result is divergent, and we need to 
include counter-terms and conditions to renormalize the effective potential. We obtain a finite effective potential, that in the limit of linear ED, it is reduced 
to the Coleman-Weinberg potential. 

%Thereby, we apply this result to Callan-Symanzik for renormalization group, and we obtain the beta function associated with the $\lambda$-scalar quartic coupling with the corrections of non-linearity through the expansion coefficients.                                
%

%
All these results were applied in the modified Maxwell electrodynamics (ModMax ED). We have showed that the quantization of the ModMax ED is achieved for a 
magnetic external field in which the vacuum energy depends only on the angle that ${\bf B}$ does with the ${\bf k}$-wave vector, and on the ModMax parameter.
The Pauli-Jordan function for ModMax is null for space-like intervals, that confirms the microcausality condition. The effective potential for the linearized 
ModMax ED also does not depends on the magnetic field, but the minimal VEV changes with the ModMax parameter, as it is illustrated in the figure (\ref{fig1}). 
As perspective of this work, the application of the vacuum energy (\ref{E0omega1}) to the Casimir Effect is an important point of investigation. The definition 
of the generating functional in a covariant gauge, and the inversion of the operator in the Green equation (\ref{EqDelta}) 
is a challenge in the path integral quantization. 
%For end, we have used the result of the beta function in usual scalar QED in (\ref{betae}) and in (\ref{betalambdae}). 
%For a more complete result, the beta function (\ref{betae}) must be corrected by the new correlation function in (\ref{ZJiJj}) through the one loop diagrams, although 
%the order in $e^{3}$ is maintained in the final result. 
All these motivations bring together questions that can be answered in a forthcoming project.    

\section*{Acknowledgements}

This work was supported by the Conselho Nacional de Desenvolvimento Cient\'ifico e Tecnol\'ogico (CNPq) under grant 305743/2026-0 (GM) (M. J. Neves).

This work is dedicated to the Professor Everton Murilo Carvalho de Abreu ({\it in memoriam}).

%%%%%%%%%%%%%%%%%%%%%%%%%%%%%%%%%%%%%%%%%%%%%%%%%%%%%%%%%%%%%%%

\appendix
\section{Integrals in $D$-dimensions of the effective potential}
In this appendix, we show in more details the results of the integrals (\ref{V1intD}). We write the regularized one loop contribution to the 
effective potential as 
\begin{eqnarray}
V_{1}(\phi_c,D)=I_{1}(\phi_c,D)+I_{2}(\phi_c,D)+I_{3}(\phi_c,D)+I_{4}(\phi_c,D) \; ,
\end{eqnarray}
where
\begin{eqnarray}
I_{i}(\phi_c,D) &=& -\frac{i}{2}\int\frac{d^Dk }{(2\pi)^{D}} \, \ln\left[\, 1-\frac{\left(\lambda_{i}/6\right) \Lambda^{2-D/2} \,\phi_c^2}{k^2-\mu^2} \, \right] \; ,
\label{Ii}
\\
I_{3}(\phi_c,D) &=& -\frac{i}{2}\int\frac{d^Dk }{(2\pi)^{D}} \, \ln\left[\, 1-\frac{ g^2 \, \Lambda^{2-D/2} \,\phi_c^2}{k^2} \, \right] 
\label{I3g}
\\
I_{4}(\phi_c,D) &=& -\frac{i}{2}\int\frac{d^Dk }{(2\pi)^{D}} \ln\left[ \frac{ (k^2-g^2 \Lambda^{2-D/2} \, \phi_c^2)^2+(k^2-g^2 \Lambda^{2-D/2} \,\phi_{c}^2)({\bf u}^2+{\bf w}^2)+{\bf u}^2\,{\bf w}^2 }{ (k^2)^2+k^2 \, ({\bf u}^2+{\bf w}^2) + {\bf u}^2 \, {\bf w}^2 } \, \right]
 , \hspace{0.9cm}
\label{I3}
\end{eqnarray}
with $\lambda_{1}=\lambda$ for $I_{1}$, and with $\lambda_{2}=3\lambda$ for $I_{2}$. The first integral has the Log-function that can be written as
\begin{eqnarray}\label{IdLog}
\ln\left[\, 1-\frac{ (\lambda_{i}/6) \, \Lambda^{2-D/2} \,\phi_c^2}{k^2-\mu^2} \, \right] =\ln\left(k^2-\mu^2-(\lambda_{i}/6) \, \Lambda^{2-D/2} \,\phi_c^2\right)
-\ln\left(k^2-\mu^2\right)
\nonumber \\
\left.
=\frac{\partial}{\partial\alpha}\left[ \, \frac{1}{\left(k^2-\mu^2 \right)^{\alpha}} 
- \frac{1}{\left(k^2-\mu^2-(\lambda_{i}/6) \, \Lambda^{2-D/2} \,\phi_c^2 \right)^{\alpha}} \, \right] \right|_{\alpha=0} \; ,
\end{eqnarray}
that substituting in (\ref{Ii}), we have the expression
\begin{eqnarray}
\left.
I_{i}(\phi_c,D) = -\frac{i}{2} \frac{\partial}{\partial\alpha} \int\frac{d^Dk }{(2\pi)^{D}} \left[ \, \frac{1}{\left(k^2-\mu^2 \right)^{\alpha}} 
- \frac{1}{\left(k^2-\mu^2-(\lambda_{i}/6) \, \Lambda^{2-D/2} \,\phi_c^2 \right)^{\alpha}} \, \right] \right|_{\alpha=0} \; .
\end{eqnarray}
These integrals can be solved using the standard integrals in $D$-dimensions from QFT \cite{Peskin} :
\begin{eqnarray}\label{IntIi}
\int\frac{d^Dk }{(2\pi)^{D}} \frac{1}{\left(k^2-\Delta^2 \right)^{\alpha}}=\frac{(-1)^{\alpha}\,i}{(4\pi)^{D/2}} \frac{\Gamma(\alpha-D/2)}{\Gamma(\alpha)} \left( \frac{1}{\Delta^2} \right)^{\alpha-\frac{D}{2}} \; ,
\end{eqnarray}
and calculating the derivative of (\ref{IntIi}) in relation $\alpha$, taking the limit $\alpha \rightarrow 0$, the result is
\begin{eqnarray}
\left.
\frac{\partial}{\partial\alpha}\left[ \, \int\frac{d^Dk }{(2\pi)^{D}} \frac{1}{\left(k^2-\Delta^2 \right)^{\alpha}} \, \right]\right|_{\alpha=0}
=\frac{i}{(4\pi)^{D/2}} \, \Gamma\left(-\frac{D}{2}\right) \, (\Delta^2)^{D/2} \; .
\end{eqnarray}
Thus, the result of (\ref{Ii}) is
\begin{eqnarray}\label{IiDresult}
I_{i}(\phi_c,D) = \frac{1}{2(4\pi)^{D/2}} \, \Gamma\left(-\frac{D}{2}\right) \left[ \, \left( \, \mu^2 \, \right)^{D/2}
-\left(\, \mu^2+ \frac{\lambda_{i}}{6} \, \Lambda^{2-D/2} \,\phi_c^2 \, \right)^{D/2} \, \right] \; , 
\end{eqnarray}
that is valid for all complex plane of $D$, expect at $\Re[D]=\{\, 2 \, , \, 4 \, , \, 6 \, \ldots \, \}$. In the approximation of 
$\lambda_{i}\,\phi_{c}^2 \gg \mu^2$ for loop integrals, the result (\ref{IiDresult}) is simplified to
\begin{eqnarray}\label{IiDresultapprox}
I_{i}(\phi_c,D) = \frac{-1}{2(4\pi)^{D/2}} \, \Gamma\left(-\frac{D}{2}\right) \, \left(\, \frac{\lambda_{i}}{6} \, \Lambda^{2-D/2} \,\phi_c^2 \, \right)^{D/2} \; .
\end{eqnarray}
Similarly, the result of (\ref{I3g}) is
\begin{eqnarray}\label{I3gDresult}
I_{3}(\phi_c,D) = \frac{-1}{2(4\pi)^{D/2}} \, \Gamma\left(-\frac{D}{2}\right) \, \left(\, g^2 \, \Lambda^{2-D/2} \,\phi_c^2 \, \right)^{D/2} \; .
\end{eqnarray} 
The integral of (\ref{I3}) contains the electromagnetic background field, that breaks the $k$-integration. 
Considering only the magnetic background field, the contractions of $F_{B}^{\mu\nu}$ and $\widetilde{F}_{B}^{\mu\nu}$ with the 
$k^{\beta}$-components are simplified as ${\bf u}^2=d_{B}F_{Bi\alpha}\,F_{B}^{\,\,\,i\beta}\,k^{\alpha}\,k_{\beta} = d_{B}\,({\bf B}\times{\bf k})^2$ 
and ${\bf w}^2=d_{E}\widetilde{F}_{Bi\alpha}\,\widetilde{F}_{B}^{\,\,\,i\beta}\,k^{\alpha}\,k_{\beta} = -d_{E} \, ({\bf B} \cdot {\bf k})^{2}$.
For the calculus of the $k$-integration in $D$-dimensions, we use the volume element in spherical coordinates 
$d^{D}k=dk_{0}\,dk \, k^{D-2} \, d\varphi \, \prod_{n=1}^{D-3} \sin^n\theta_{n} \, d\theta_{n}$, where the spatial $k$-components is on the radial direction.  
Assuming the magnetic field on the ${\cal Z}$-direction, with ${\bf B}=B\, \hat{{\bf z}}$, we have 
${\bf u}^2=B^2\,k^2\sin\theta$ and ${\bf w}^2 = -B^2\,k^2\,\cos^2\theta$, where $\theta$ is the angle that the radial direction does with the ${\cal Z}$-axis. Thereby, using the similar identity to (\ref{IdLog}), the $I_{3}$-integral is written as 
\begin{eqnarray}\label{IntI3alpha}
&&
\left.
I_{4}(\phi_{c},D)=\frac{-i}{2(2\pi)^{D-1}} \, \frac{\partial}{\partial\alpha} \prod_{n=1}^{D-3} \int_{0}^{\pi} (\sin\theta)^{n} \, d\theta  
\, \int_{0}^{\infty} dk \, k^{D-2} 
\right.
\nonumber \\
&&
\left.
\times 
\int_{-\infty}^{\infty} dk_{0}
\left\{ \, \frac{1}{\left[ \, (k_{0}^2-k^2)^2+ (k_{0}^2-k^2) \, k^2\,C_{1}-C_{2}\,k^4 \, \right]^{\alpha}} 
\right.
\right.
\nonumber \\
&&
\left.
\left.
-\frac{1}{\left[ \, (k_{0}^2-k^2-g^{2}\Lambda^{2-D/2}\,\phi_c^2)^2+ (k_{0}^2-k^2-g^{2}\Lambda^{2-D/2}\,\phi_c^2) \, k^2\,C_{1}-C_{2}\,k^4 \, \right]^{\alpha}} \, \right\} \right|_{\alpha=0} ,
\end{eqnarray}
where $C_1=B^2\,(d_{B}\,\sin^2\theta-d_{E}\cos^2\theta)$ and $C_2=d_{B}\,d_{E}\,B^4\,\sin^2\theta\,\cos^2\theta$.

The first integral in (\ref{IntI3alpha}) is null due to property for $D$-dimensions :
\begin{eqnarray}
\int \frac{d^{D}k}{(2\pi)^{D}} \frac{1}{(k^2)^{\alpha}} = 0 \; .
\end{eqnarray}  
The second integral in (\ref{IntI3alpha}) is calculated in $k^{0}$, and after in $k$-radial to reduce it to 
angular integral in $\theta$
\begin{eqnarray}\label{I3inttheta}
\left.
I_{4}(\phi_{c},D)=\frac{i}{4\,(2\pi)^{D-1}} \, \frac{\partial}{\partial\alpha} \left[ \, (-1)^{1/2-\alpha} \, \frac{\Gamma(\alpha-D/2)}{\Gamma(\alpha)} \, \left( \, g^2\,\Lambda^{2-D/2}\,\phi_{c}^2 \, \right)^{\frac{D}{2}-\alpha} \, \right] \right|_{\alpha=0}   
\nonumber \\
\times \, 
\Gamma\left( \frac{D-1}{2} \right) \,\prod_{n=1}^{D-3} \int_{0}^{\pi} \, d\theta  \, \frac{(\sin\theta)^n}{\left[ \, 1+ B^2\left(d_{B}\,\sin^2\theta
-d_{E}\,\cos^2\theta\right) \, \right]^{(D-1)/2}} \; .
\end{eqnarray}
After the calculus of the $\alpha$-derivative, and taking the limit $\alpha \rightarrow 0$, the $\theta$-integral with the product 
yields the result :
\begin{equation}\label{I3ResultD}
I_{4}(\phi_{c},D)=\frac{-1}{4\,(4\pi)^{D/2}} \, \Gamma\left(-\frac{D}{2}\right) \, \frac{\left( \, g^2\,\Lambda^{2-D/2}\,\phi_{c}^2 \, \right)^{D/2}}{\left[\,\left(1+d_{B}B^2\right)\left(1-d_{E}B^2\right)\,\right]^{(D-1)/2}} \; ,
\end{equation}
%
%\begin{eqnarray}\label{I3ResultD}
%&&
%I_{3}(\phi_{c},D)=-\frac{3}{4\,(4\pi)^{D/2}} \, \Gamma\left(-\frac{D}{2}\right) \, \left( \, g^2\,\Lambda^{2-D/2}\,\phi_{c}^2 \, \right)^{D/2}
%\nonumber \\
%&&
%\hspace{-0.5cm}
%\times 
%\left[ \, \frac{\sqrt{\pi}}{2}\frac{(1+d_{B}B^2)^{1-D/2}}{B \,\sqrt{d_E+d_B}} %\frac{\Gamma\left(\frac{D}{2}\right)\Gamma\left(\frac{1}{2}-D\right)}{\Gamma\left(2-\frac{D}{2}\right)\Gamma\left(\frac{D}{2}-\frac{1}{2}\right)}
%+ \frac{\sqrt{\pi}}{2}\frac{(1+d_{B}B^2)^{1-D/2}}{B \,\sqrt{d_E+d_B}} \frac{\Gamma\left(\frac{3}{2}-\frac{D}{2}\right)}{\Gamma\left(2-\frac{D}{2}\right)} 
%\right.
%\nonumber \\
%&&
%\hspace{-0.5cm}
%\left.
%+\frac{(1-d_{E}B^2)^{(3-D)/2}}{(D-3)\sqrt{1+d_{B}B^2}} \, {}_2F_1\left(1,2-\frac{D}{2},\frac{5-D}{2};\frac{1-d_{E}B^2}{1+d_{B}B^2}\right)
%\right.
%\nonumber \\
%&&
%\left.
%-\frac{(1-d_{E}B^2)^{(3-D)/2}}{D-2} \, {}_2F_1\left( 1, \frac{D-1}{2}, \frac{D}{2};\frac{1+d_{B}B^2}{1-d_{E}B^2}\right)
%\, \right] \; ,
%\end{eqnarray}
%
in which the analytical extension of the Gamma function in the $D$-complex plane excludes the values $\Re[D]=\{\, 2 \, , \, 4 \, , \, 6 \, \ldots \, \}$.
For more details on the integral and the product in (\ref{I3inttheta}), see the ref. \cite{Gradstheyn}). Summing (\ref{I3ResultD}) with 
(\ref{IiDresultapprox}) and (\ref{I3gDresult}), we obtain (\ref{V1intDresult}).

%**************

%
%\begin{eqnarray}
%I_{3}(\phi_c,D) = \frac{-3i}{2(2\pi)^{D-1}} \prod_{n=2}^{D-3} \int_{0}^{\pi} \sin^n\theta_{n} \, d\theta_{n} \int_{0}^{\pi} \sin\theta \, d\theta 
%\, \int_{0}^{\infty} dk \, k^{D-2} 
%\nonumber \\
%\times \int_{-\infty}^{\infty} dk_{0} \,
%\ln\left[ \, 1-\frac{ e^2 \, \Lambda^{2-D/2} \, \phi_c^2 }{k_{0}^2-k^2-d_{B}\,B^2\,k^2\,\sin^2\theta/3+d_{E}\,B^2\,k^2\cos^2\theta/3 } \, \right] \; .
%\end{eqnarray}
%
%Making the $k_{0}$-integration, we obtain
%
%\begin{eqnarray}
%I_{3}(\phi_c,D) = \frac{-3i}{2(2\pi)^{D-1}} \frac{\pi^{D/2-3/2}}{2\,\Gamma(\frac{D}{2}-\frac{1}{2})} \int_{0}^{\pi} \sin\theta \, d\theta 
%\, \int_{0}^{\infty} dk \, k^{D-2} 
%\nonumber \\
%2\pi\,i \left[ \, \sqrt{ \, k^2+e^2 \, \Lambda^{2-D/2} \, \phi_c^2+d_{B}\,B^2\,k^2\,\sin^2\theta/3-d_{E}\,B^2\,k^2\cos^2\theta/3 \, } 
%\right.
%\nonumber \\
%\left.
%- \sqrt{ \, k^2+d_{B}\,B^2\,k^2\,\sin^2\theta/3-d_{E}\,B^2\,k^2\cos^2\theta/3 \, } \, \right] \; .
%\end{eqnarray}
%

%
%\section*{Acknowledgement}
%

%

%************************************

%
\end{document}